\documentclass[useAMS,usenatbib]{mn2e}
\usepackage{epsfig}
\usepackage{times}

\def\cm3{cm$^{-3}$}
\def\kms{km~s$^{-1}$}

\def\lsun{L$_{\odot}$}

\def\msun{M$_{\odot}$}

\def\beq{\begin{equation}}
\def\eeq{\end{equation}}

\title[Time-dependence Effects in Photospheric-Phase Type II Supernova Spectra]
{Time-dependence Effects in Photospheric-Phase Type II Supernova Spectra}
\author[Luc Dessart and D. John Hillier]{Luc Dessart$^{1}$\thanks{E-mail: luc@as.arizona.edu}
and D. John Hillier$^{2}$\\
$^{1}$ Department of Astronomy and Steward Observatory,
                 The University of Arizona, Tucson, AZ \ 85721\\
$^{2}$ Department of Physics and Astronomy, University of Pittsburgh \\
}

\begin{document}

\date{Accepted . Received }

\pagerange{\pageref{firstpage}--\pageref{lastpage}} \pubyear{2007}

\maketitle

\label{firstpage}

\begin{abstract}

Spectroscopic modeling of Type II supernovae (SNe) generally assumes
steady-state. Following the recent suggestion of Utrobin \& Chugai, but
using the 1D non-LTE line-blanketed model atmosphere code
CMFGEN, we investigate the effects of including time-dependent terms, which are
generally neglected, that appear in the statistical and radiative equilibrium equations.
We base our discussion on the ejecta properties and the spectroscopic signatures obtained
from time-dependent simulations, investigating different ejecta configurations
(slow, standard, and fast), and covering their evolution from one day to six weeks
after shock breakout.
Compared to equivalent steady-state models, our time-dependent models produce SN ejecta
that are systematically over-ionized, affecting helium
at one week after explosion, but ultimately affecting all ions after a few weeks.
While the continuum remains essentially
unchanged, time-dependence effects on observed spectral lines are large.
At the recombination epoch, H{\sc i} lines and Na{\sc i}\,D are considerably
stronger and broader than in equivalent steady-state models, while Ca{\sc ii}\,8500\AA\ is weakened.
If time dependence is allowed for, the He{\sc i} lines at 5875\AA\ and
10830\AA\ appear $\sim$3 times stronger at one week, and He{\sc i}\,10830\AA\
persists as a blue-shifted absorption feature even at 6 weeks after explosion.
Time dependence operates through the energy gain
from changes in ionization and excitation, and, perhaps more universally across SN types,
from the competition between recombination and expansion, which in-turn, can be affected by
optical-depth effects.
Our time-dependent models compare well with observations of
the low-luminosity low-velocity SN\,1999br and the more standard
SN\,1999em, reproducing the H$\alpha$ line strength at the recombination epoch,
and without the need for setting unphysical requirements on the magnitude of nickel mixing.

\end{abstract}


\section{Introduction}

Because of radiative cooling and the fast expansion of the exploding mantle
of the progenitor massive star, photospheric-phase Type II SN spectra evolve rapidly.
At shock breakout the spectral energy distribution (SED) peaks in the far-UV and X-rays.
Subsequently the SED centers in the UV, then in the optical, and later in
the infrared until the object fades from view.
From a radiative transfer perspective, the cooling
induces a recombination to lower ionization stages that impose a strong blanketing
effect on the energy distribution. A few days after explosion, the ejecta radiate
a UV-dominated SED, with He{\sc ii}\,4686\AA\ (only observed at a few days after shock breakout;
Dessart et al. 2007), Balmer/Paschen lines of Hydrogen, He{\sc i}\,5875\AA, He{\sc i}\,10830\AA,
some multiplets of O{\sc ii} and N{\sc ii} multiplets at \,4600\AA\ (see Dessart \& Hillier 2006a;
Baron et al. 2007), N{\sc ii}\,5400\AA,
and a few isolated resonance lines such as Mg{\sc ii}\,2802\AA\ and Al{\sc iii}\,1859\AA.
As the ejecta continue to cool, hydrogen eventually recombines, with a contemporaneous enhancement
in line-blanketing due to the switch from Fe{\sc iii} to Fe{\sc ii}. In Dessart \& Hillier (2006)
and Dessart et al. (2007), we successfully modeled these epochs for SNe 1999em,
2005cs, and 2006bp, using steady-state non-LTE CMFGEN models and a power-law density distribution
with an exponent of ten (the immediate post-breakout phase requires higher values for this exponent).
These spectroscopic analyses deliberately focused on the first $\sim$45 days after shock breakout,
since, beyond that time, we encountered severe difficulties in reproducing
the hydrogen lines. Specifically, at late times, synthetic line profiles were sizably narrower
than observed, suggesting that line formation regions predicted by CMFGEN
were confined to velocities/radii that were too small. Despite a continued success with reproducing
the overall continuum energy distribution, CMFGEN failed to reproduce important line profiles,
whenever hydrogen started recombining at and above the photosphere.

   The H$\alpha$ problem has not been clearly emphasized in the literature, where
one can see a great disparity in model atmosphere assumptions and agreement between
theoretical predictions and observations, but no direct link to physical/numerical issues.
In the eighties and early nineties, the recognized importance of non-LTE effects
confronted the strong limitations of computer technology, so that only a few species
were treated in non-LTE (i.e., usually hydrogen and helium), while the metals
responsible for line blanketing were treated in LTE.
\citet{Eastman_Kirshner_1989} followed this approach to model the
first ten days of SN 1987A, and did not encounter obvious difficulties with
hydrogen lines. \citet{Schmutz_etal_1990}, using an approximate non-LTE technique,
had, on the contrary, great
difficulty reproducing any of the Balmer lines, suggesting clumping as the culprit.
\citet{Hoeflich_1988} reproduced the SN 1987A spectral evolution and the hydrogen lines,
over many months, using a large ``turbulent'' velocity.
In the more sophisticated non-LTE CMFGEN \citep{HM_98,DH_05a}
models presented in \citet{DH_06a}, we found that decreasing the
turbulent velocity weakened line-blanketing
effects and increased, although only modestly, the strength of hydrogen lines.
However, beyond 40 days after
explosion, this tuning had no longer any important influence on the hydrogen lines.

The non-LTE model atmosphere code PHOENIX predicts strong Balmer lines at the
hydrogen recombination epoch \citep{Mitchell_etal_2001,Baron_etal_2003},
but this may stem from their adoption of
non-thermal ionization/excitation due to $^{56}$Ni at the photosphere, sometimes
just a few days after explosion and in mass shells moving at $\ge$10000\,\kms
in SN 1987A \citep{Mitchell_etal_2001} or $\sim$9000\,\kms in SN 1993W \citep{Baron_etal_2003}.
Hydrodynamical simulations of core-collapse SNe predict that $^{56}$Ni
has velocities of at most $\sim$4000\,\kms, the nickel fingers being strongly
decelerated at the H/He interface \citep{Fryxell_etal_1991,Kifonidis_etal_2000, Kifonidis_etal_2003}.
The magnitude of this disagreement extends far beyond the uncertainties of explosion
models and suggests a genuine incompatibility.

  Interestingly, H{\sc i} lines remain strong for months in all Type II SNe, in objects
as diverse as the ``peculiar'' SN 1987A, the ``plateau'' SN 1999em,
and the ``low-luminosity'' SN 1999br.
CMFGEN models computed for these objects were unable to reproduce Balmer lines
after $\sim$4 days in SN 1987A, $\sim$40 days in SN 1999em, and $\sim$20 days in SN 1999br,
all coincident with hydrogen recombination in the ejecta.
These three SNe have very different inferred ejecta properties and observed light curves.
SN 1999br even synthesized an order of magnitude less $^{56}$Ni than average
\citep{Pastorello_etal_2004} for the Type II class.
The only common property between these objects, which is connected to the H$\alpha$ problem,
is the recombination of the ejecta to a lower ionization state at the corresponding epoch.

Recently, \citet[UC05]{UC_05} proposed that the effect of time-dependence, and
the energy associated with changes in ionization/excitation, lead to a strong
H$\alpha$ line profile in SN 1987A during the recombination epoch.
They also found that barium lines were affected, and that, with their more
consistent approach, Ba{\sc ii}\,6142\AA\ could be fitted using the LMC metallicity value.
The steady-state models of \citet{Mazzali_etal_1992} supported instead an abundance enhancement
of five.

Time dependence has been invoked in the past by \citet{Fransson_Kozma_1993} to explain the
late-time light curve of SN 1987A, and the theoretical study of \citet{Pinto_Eastman_2000a,
Pinto_Eastman_2000b} showed that time dependence {\it in the radiation field} had a
critical impact on the radiative transfer in Type Ias.
\citet{Pinto_Eastman_2000a, Pinto_Eastman_2000b}, however, treat the material in
LTE, i.e., do not solve the rate equations,
focusing instead on the time-dependent {\it diffusion} of photons through an optically
thick Type Ia SN ejecta. Similarly, \citet{Kasen_etal_2006} neglect explicit time
dependence in the rate equations, computing the ionization and excitation state of the
medium in LTE. Thus, although there is at present growing interest in accounting for time
dependence in the radiation field, the often-used expedient of LTE, to maintain low CPU costs, has forced the neglect of both non-LTE and time-dependence in the level populations.
One exception to these time-dependent LTE approaches is the work of \citet{Hoeflich_2003}, who
treats time-dependence in the rate equations but, to our knowledge,
has not discussed the associated effects on Type II SN spectra.
In the present study, we investigate thoroughly the effects of time-dependence
in the rate equations and their impact on inferred ejecta properties. Note that
time dependence in the radiation field is accounted for by adjusting the base luminosity
so that the emergent synthetic flux matches the observed flux, using the bolometric-light
evolution of SN 1999em as a guide \citep{DH_06a}.

   Here, we report the salient features of several {\it time-dependent} non-LTE
CMFGEN simulations, covering the early evolution for a range of Type II SN ejecta.
We confirm the results of UC05 that time-dependence induces an over-ionization of
the recombining ejecta and that it solves the H$\alpha$ problem.
However unlike UC05, we self-consistently solve the radiation
transfer equation --- the coupling between the level populations and the radiation field
is calculated and fully allowed for. Our more comprehensive study of the full spectrum
further predicts that all lines, not just those of hydrogen, are significantly affected.
The ionization of the ejecta and its evolution are so strongly modified that we
predict even He{\sc i}\,10830\AA\ many weeks after explosion.

In the next section, we present out treatment of time-dependence in CMFGEN, which focuses
here on the terms appearing in the statistical and radiative equilibrium
equations (an appendix also provides further details).
In \S2.5, we present the various model calculations performed to illustrate
our discussion. In \S3, we present our results, discussing the effects of time dependence on
the ejecta properties as well as the associated spectroscopic signatures.
We then present in \S4 a comparison with a few representative observations, focusing on
the well observed Type II Plateau SN 1999em and the low luminosity Type II SN 1999br.
In \S5, we discuss the implications of such time-dependent effects on our understanding
and on our modeling of photospheric-phase Type II SN spectra, before giving our conclusions
in \S6.
Note that the salient features and key results presented here are also available in
a concise, ``letter'', format in \citet{DH_06b}, to which we refer the hurried
reader.

\section{Treatment of time dependence}

 \subsection{Statistical Equilibrium Equations}

 The general form of the time dependent statistical equilibrium equation
is

\begin{equation}
  {\partial  n_i \over \partial t} + \nabla.(n_i v) =
          \sum_j \left(n_j R_{ji}  - n_i R_{ij}\right)\,,
 \end{equation}

\noindent
where $v$ is the velocity, $n_i$ is the population of state $i$, and $R_{ij}$ is the
transition rate from state $i$ to state $j$. The forms
of $ R_{ij}$ and  $R_{ji}$ are standard, and can be found, for example, in Mihalas
(1978). For simplicity we have not distinguished between the different
processes (collisional, bound-free, and bound-bound) entering these
rates.

In massive stars, the winds can be approximated as stationary. In such cases
we consider a Eulerian formulation, and take $\partial n/\partial t$ to
be zero. The $\nabla.(n_iv)$ term is the advection term, and is
usually neglected, however it can become
important in the outer regions of the wind, since the
ratio of the recombination time, $1/{\alpha n_e}$, to the flow time, $r/v$,
scales as the radius, $r$ (here, $\alpha$ is the recombination coefficient).

Supernovae are obviously time dependent, and in this case it makes sense to
use a Lagrangean formalism. Using the continuity equation

\begin{equation}
{D \rho \over Dt} + \rho (\nabla.v) =0\,,
\end{equation}

\noindent
the statistical equilibrium equations become

\begin{equation}
 \rho {D n_i/\rho \over Dt} =
          \sum_j \left(n_j R_{ji}  - n_i R_{ij}\right)\,,
\end{equation}

\noindent
where $D/Dt$ is the comoving derivative. For a Hubble flow,

\begin{equation}
 {1 \over r^3} {D r^3 n_i  \over Dt} =
           \sum_j \left(n_j R_{ji}  - n_i R_{ij}\right)\,,
 \end{equation}

\noindent
which is the form we used in these calculations.

To solve these equations we use implicit first order differencing
in the Lagrangean frame. Thus the equations to be solved become

\begin{equation}
{(n_i)_k- (n_i)_{k-1}(r_{k-1}/r_k)^3  \over \Delta t} =
          \sum_j \left( n_j R_{ji}  - n_i R_{ij} \right)_k
\end{equation}

\noindent
where $k-1$, $k$ refer to values at consecutive time steps
but with the same comoving coordinate.
For a Hubble flow, we can simply use the velocity as the Lagrangean
co-ordinate. The solution of these equations proceeds in
an identical fashion to that used to solve the steady-state
statistical equilibrium equations (Hillier 1987;
Hillier 1990; Hillier \& Miller 1998); the only distinction is that
we have a ``source'' term that comes from the populations
at an earlier time step.

\subsection{Energy balance}

The energy equation is

\begin{equation}
\rho {De \over Dt} - {P \over \rho} {D\rho \over Dt}= 4\pi \int (\chi_\nu J_\nu  - \eta_\nu)\,,
\end{equation}

\noindent
where $e$ is the internal energy/unit mass, $\chi_\nu$
the opacity, $\eta_\nu$ the emissivity, and $J_\nu$ the mean intensity, and
we have ignored other forms of energy deposition (such as nuclear decay).
$e$ can be written in the form

\begin{equation}
   e = e_K +e_I\,,
\end{equation}

\noindent
where

\begin{equation}
   e_K={3 kT(n+n_e) \over 2 \mu m n }\,,
 \end{equation}

\noindent
and

\begin{equation}
e_I=  \sum_i {n_i E_i \over \mu m n}\,.
\end{equation}

\noindent
In the above, $n$ is the total particle density (excluding electrons),
$n_e$ is the electron density, and $E_i$ is the total energy (excitation and ionization)
of  state $i$. This equation is solved implicitly, via linearization,
and is no more difficult to treat than the regular constraint of
radiative equilibrium.

\subsection{Are the time-dependent terms important?}
\label{sect_ddt_tau}

Below we examine when the time dependent terms in the statistical equilibrium equations will play an important role in determining the radiative transfer in SN envelopes.
To do this, we need to compare the time it takes to replenish the population of individual
levels with the flow time. As noted previously, the flow-time scale is simply $r/v$,
and for a Hubble flow this is $t_{\hbox{exp}}$, where $t_{\hbox{exp}}$ is the time since the explosion.

We will only examine the ionization equilibrium in detail, since it is the ionization equilibrium
that will be most affected by the time-dependent terms --  statistical equilibrium of excited levels
will generally be satisfied because the radiatives rates tend to be large
($A> 10^6$\,s$^{-1}$), and the transitions have low (or moderate) optical depths.

The simplest recombination timescale can be found by estimating the
time it  takes, at fixed electron density, for all ions to recombine.
Thus the recombination timescale for hydrogen, $t_{\hbox{rec}}$ is simply given by

\begin{equation}
 t_{\hbox{rec}} = {n({\rm H}^+) \over n({\rm H}^+) n_e \alpha} =
{1 \over n_e \alpha}\,.
\end{equation}

Using $\alpha=\alpha_B=2.59 \times 10^{-13}$\,cm$^{3}$\,s$^{-1}$ at 10,000K
\citep{Osterbrock_1974}, gives

\begin{equation}
t_{\hbox{rec}} = {4.47 \times 10^7 \over n_e} \,\, {\rm days.}
\end{equation}

\noindent
We use $\alpha_B$ since in SN conditions  every direct recombination
to the ground state will generally be followed by
an ionization. The recombination time scale for He{\sc i} is very
similar to that of H{\sc i}.
Since the comoving density is $\propto  1/t^3$, the ratio of
recombination time to the flow time, at a given velocity,
scales as $t^2$. Thus the ionization must eventually become
frozen, and first at high velocities.

The previous discussion refers to the case when
we can ignore optical depth effects and the existence of metastable
states.  When optical depth effects are important, and/or there
exist low lying metastable states, it is more difficult to ascertain
the relevant recombination time scale. In such cases, it is the
net flow of electrons to the ground state that will set the ionization
equilibrium.

As an illustration consider a two level atom with continuum. The
relevant equations are

\begin{equation}
{Dn_1 \over Dt} =   n_2 R_{21} - n_1 R_{12}\,,
\end{equation}

\noindent
and

\begin{equation}
{Dn_2 \over Dt} =  n_e n_I \alpha - n_2 P_{2I}  + n_1 R_{12} - n_2 R_{21}  \,,
\end{equation}

\noindent
and

\begin{equation}
n_1 + n_2 +n_I= n\,\,\, .
\end{equation}

\noindent
Here $n_I$ is used to denote the ion population, $P_{2I}$ is the photoionization rate,
$R_{21}$  refers to the combined rate for processes which couple level 2 to
1 (bound-bound, collisional, two photon decay), while $R_{12}$ refers to
processes that excite an electron from level 1 to 2. We assume that the
ground state continuum is in detailed balance. For a strict two-level atom
$\alpha=\alpha_{I2}$ but for these discussions we can take $\alpha=\alpha_B$.

In the absence of the time-dependent terms, the ground state equation reduces to

\begin{equation}
n_2 R_{21} = n_1 R_{12} \,.
\end{equation}

\noindent
In the case of He{\sc i} the main decay route from the 2s\,$^1S$ state
is two photon decay (with $A=52$\,s$^{-1}$), while collisional excitation might be the main reverse
mechanism.

When we include the time-dependent terms, the equation for level 1
can reduce to

\begin{equation}
{D n_1 \over Dt} =   n_2 R_{21}\,,
\end{equation}

\noindent
and that for level 2 becomes

\begin{equation}
n_2 P_{2I} =  n_e n_I  \alpha\,,
\end{equation}

\noindent
since the decay term, for a metastable level or for a level decaying through an optically
thick transition, can be small. The effective recombination
time scale, $t_{\hbox{eff}}$,  is (since $Dn_1/Dt=-DX^+/Dt$) thus

\begin{eqnarray}
t_{\hbox{eff}} = & n_I \over Dn_1/Dt   \\
                     = & n_I  \over n_2 R_{21} \\
                     = & n_I \over n_e  n_I  \alpha R_{21}/P_{2I} \\
                     = & t_{\hbox{rec}} {P_{2I} \over R_{21}} \,.
\end{eqnarray}

Thus, if the photoionization rate $P_{21}$ is much greater than the decay rate
$R_{21}$, the effective recombination time scale will be considerably
longer than the classic recombination time scale.

The expression for $P_{2I}$ is

\begin{equation}
P_{2I} = \int_{\nu_o}^\infty  \sigma \left( {4\pi \over h \nu} \right)  J_\nu \, d\nu \,,
\end{equation}

\noindent
where $\sigma$ is the photoionization cross-section and $h$ is
Planck's constant. Assuming $J_\nu=  WB_\nu$ (a diluted blackbody),
$\sigma=\sigma_o (\nu_o / \nu)^2$, and $h\nu_o/kT >>1$, we have

\begin{eqnarray*}
P_{2I} = & 5.8 \times 10^6 W \left( {\sigma_o \over 10^{-18}}\right)
\left( { \nu_o \over 10^{15} {\rm Hz}} \right)^2  \\
& \left( {T \over 10^4 {\rm K} } \right) \exp(-h\nu_o/kT) \,\,\, {\rm s}^{-1} \,.
\end{eqnarray*}

\noindent
For He\,{\sc i} 2s\,$^1S$, we have $A($two phot$)=52$\,s$^{-1}$,
$\nu_o=0.960 \times 10^{15}$\,Hz, $\sigma_o=8.90 \times 10^{-18}$\,cm$^2$,
and, assuming $W=1$, and  $T=10^4$\,K, we find
$$t_{\hbox{eff}}=9.2 \times 10^3 \, t_{\hbox{rec}} \,\,.$$
Thus, it is apparent that the effective recombination time scale is much longer than the
classic recombination time scale, and could easily be comparable
to the flow time scale.

We appreciate that the true situation is more complex
than that discussed above. For He{\sc i} the 2s\,$^1S$
and 2p\,$^3S$ state are both metastable, and there is
a collisonal coupling between the singlet and triplet terms. The 2s\,$^1S$
state can decay directly to the ground state, for example,
by two-photon decay, by collisional de-excitation, and
by collisional excitation to the 2p\,$^1$P$^{\rm o}$ state followed
by a radiative decay. Further, the effective
recombination rate will substantially exceed the
classic recombination time scale only if the resonance line
is optically thick. Despite these caveats, it is apparent that
the time-dependent terms might be important even at the photosphere, and must
be included. A simple comparison between the flow time
scale and the classic recombination time scale can greatly
underestimate the importance of these time-dependent terms.

In the Appendix we examine, in detail, the various processes
that influence the ionization of helium at a  representative
depth, and in a typical model, and where the flow time is significantly greater
than the classic recombination time. In that appendix we show
that the combination of optical depth effects, and the existence of
metastable levels, leads the time-dependent terms to cause a delay in the
recombination of He$^+$ to neutral He.

For H{\sc i}, the 2s state is metastable, but in atmospheric
calculations (as is done here) it is usually assumed to be
in LTE with respect to the 2p state (i.e., the 2s and 2p states have the same departure
coefficients). Thus the relevant time scale
is set by radiative decay from level 2p to 1s, and, hence, by the
Ly$\alpha$ optical depth, as was previously noted by
UC05. Other processes or configurations (e.g., overlapping lines,
continuum opacity) can also be important since they can destroy
a Ly$\alpha$  photon, facilitating a ``real'' recombination to the ground state.

Finally we stress, as apparent from the above discussion, that the time-dependent term
does not directly change the ionization equilibrium (unlike
the situation at low electron densities). Rather, it prevents the
decay of electrons from the $n=2$ states (for H\,{\sc i} and He\,{\sc i})
to the ground state. The ionization of H\,{\sc i} and He\,{\sc i} can then be
maintained by ionizations from these excited states.

\subsection{Radiative Transfer}

\begin{table*}
  \centering
  \begin{minipage}{180mm}
  \caption{CMFGEN ejecta characteristics. From left to right, the column refers to the
model name, the initial base radius $R_0$, the base velocity $v_0$, the initial base density $\rho_0$, the total mass of the ejecta covered by the CMFGEN grid $M_{\rm ejecta}$,
the initial and final times $t_i$ and $t_f$, the initial and final
base electron-scattering optical depths $\tau_{\rm es, i}$ and $\tau_{\rm es, f}$,
assuming full and uniform ionization ($\kappa_{\rm es} = 0.30$), the initial and
final base electron-scattering optical depths  $\tau_{\rm es, i}$ and $\tau_{\rm es, f}$
computed with CMFGEN {\it including time-dependent} terms in the statistical
and radiative equilibrium equations, and, in the last column, the final base
electron-scattering optical depth computed with CMFGEN under the assumption of
{\it steady-state}. Note how the value of $\tau_{\rm es, f}$ diminishes from
the assumption of fixed and full ionization, CMFGEN with time-dependence, and
CMFGEN without time-dependence.\label{table_model_param}}
  \begin{tabular}{lcccccccccccc}
  \hline
& & $R_0$ & $v_0$ & $\rho_0$ & $M_{\rm ejecta}$ & $t_{\rm i}$ & $t_{\rm f}$
 & $\tau_{\rm es, i}$ & $\tau_{\rm es, f}$ & $\tau_{\rm es, i}$  &
$\tau_{\rm es, f}$ & $\tau_{\rm es, f}$ \\
& &  10$^{10}$\,cm  & \kms & 10$^{-10}$\,g\,cm$^{-3}$ & \msun &
\multicolumn{2}{c}{days}& \multicolumn{2}{c}{Fixed ionization}&
\multicolumn{2}{c}{CMFGEN} & CMFGEN \\
 & & \multicolumn{8}{c}{}& \multicolumn{2}{c}{D/DT}& no D/DT \\
\hline
Baseline & A & 25238 & 3471 & 4.2 & 6.17 & 8.42 & 48.68 & 3686 & 111.4 & 3640 & 45.0 & 0.9 \\
Slow     & B & 14000 & 2000 & 8.0 & 2.00 & 8.10 & 42.82 & 3886 & 143.9 & 3840 & 9.8  & 0.6 \\
Fast     & C & 25238 & 4400 & 4.2 & 6.17 & 6.63 & 35.51 & 3686 & 135.4 & 3640 & 20.7 & 0.4 \\
Early    & D & 6539  & 4350 & 4.0 & 0.10 & 1.74 & 6.76  & 904  & 60.2  & 900  & 51.9 & 51.7 \\
\hline
\hline
\end{tabular}
\end{minipage}
\end{table*}

\begin{figure*}
\epsfig{file=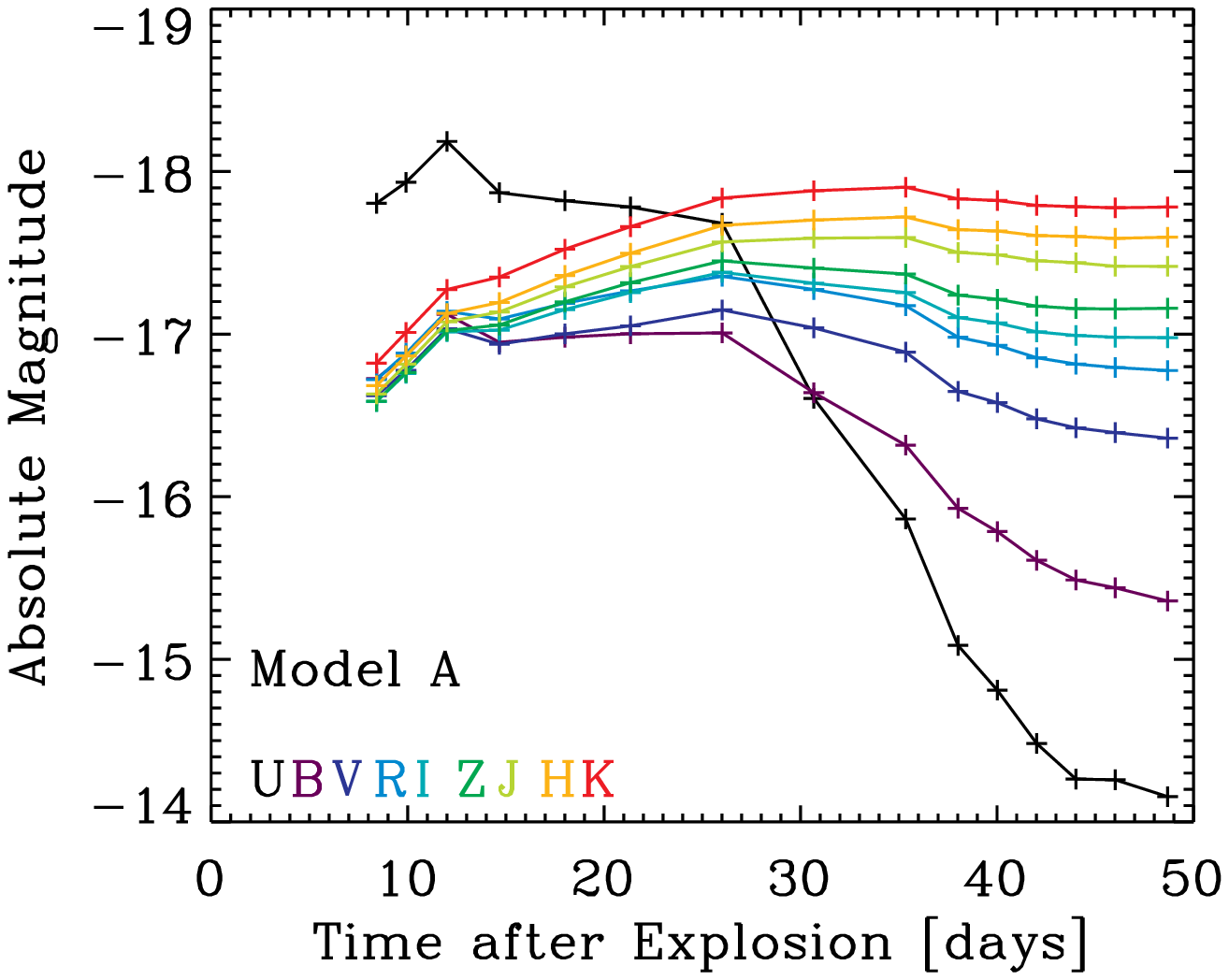,width=8cm}
\epsfig{file=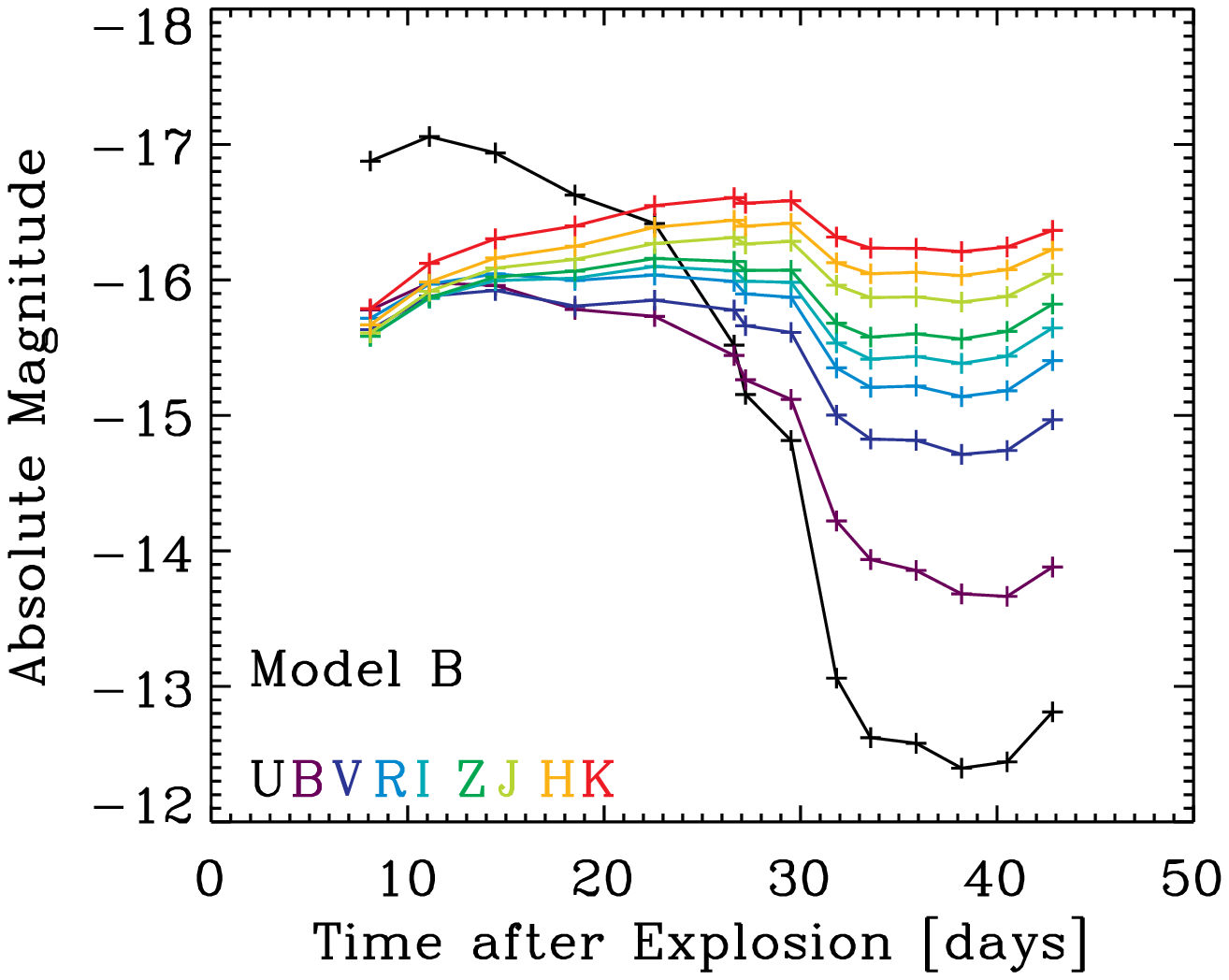,width=8cm}
\epsfig{file=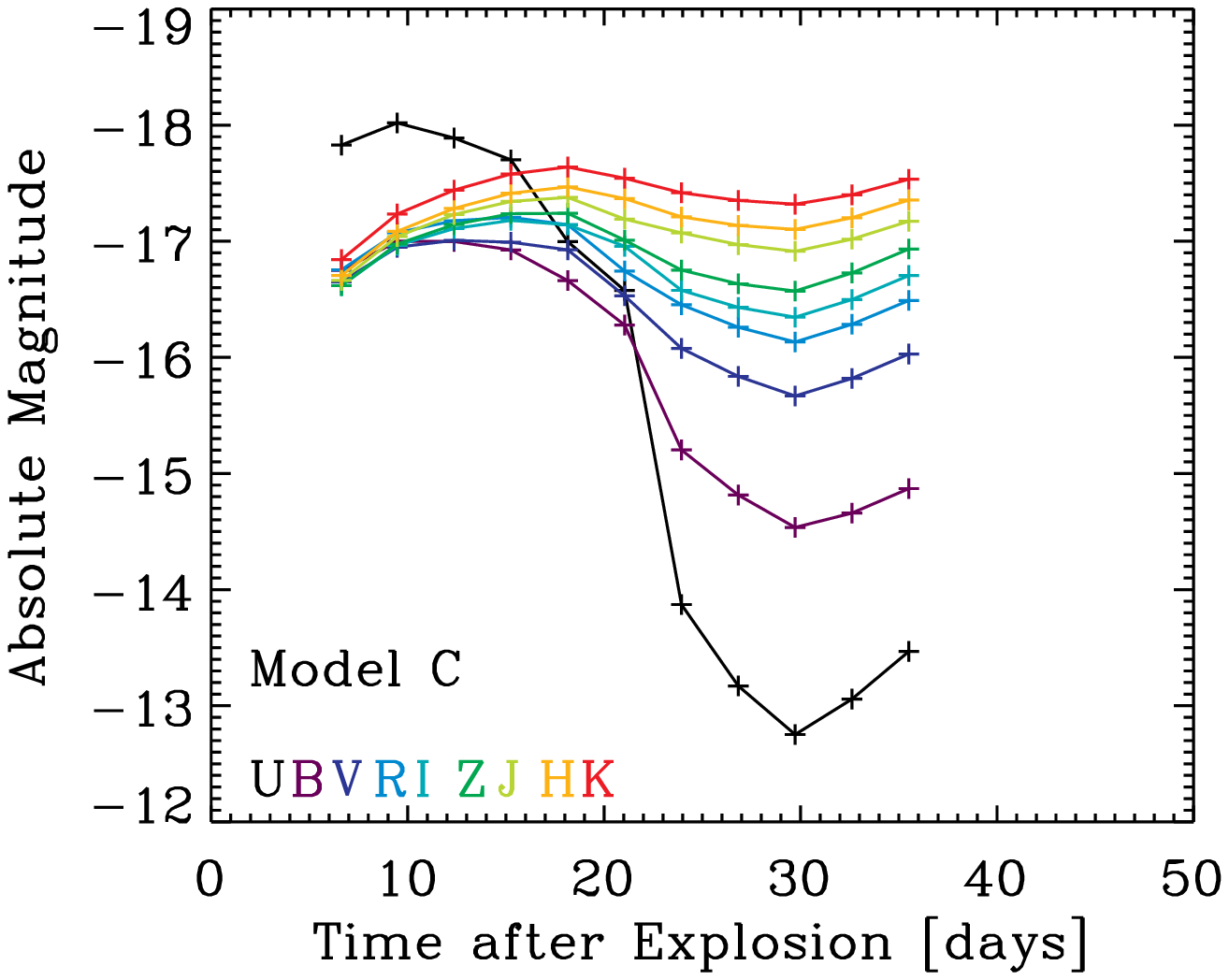,width=8cm}
\epsfig{file=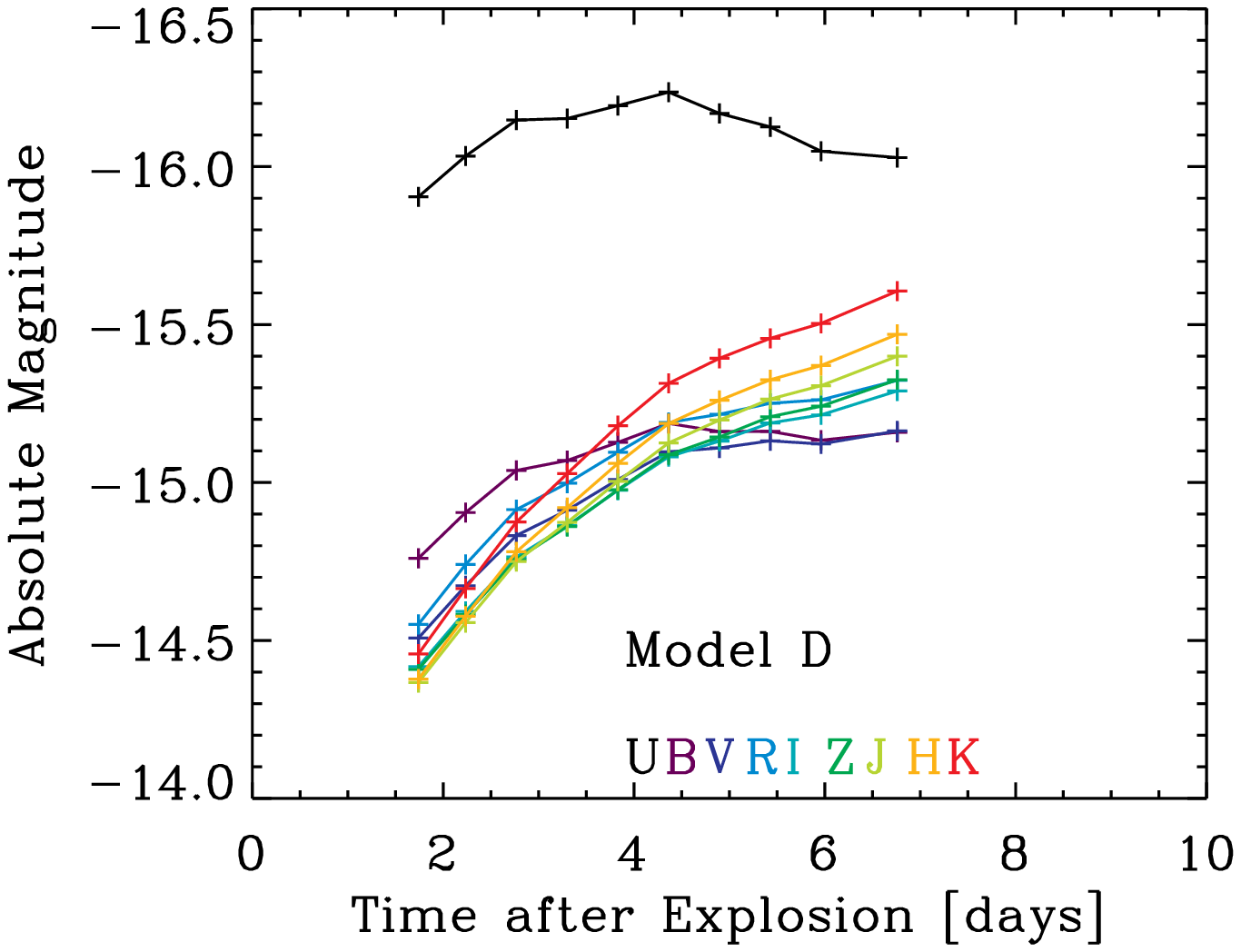,width=8cm}
\caption{Synthetic light curves for the baseline model (A), the slow (B) and the fast (C) version
of model A, and the ``early'' model (D), which covers the first week after explosion.
These models correspond to standard (models A and C) and underluminous (models B and D)
Type II SNe, and reproduce approximately the evolution of their bolometric luminosity and color,
with a bolometric luminosity that decreases and an SED that peaks further and further to the red with time.
Note that in some models, the emergent luminosity shows a small rise at the end of the
sequences, resulting from a slight inadequacy in the imposed flux at the inner boundary.
[See the electronic edition of the Journal for a color version of this figure.]
}
\label{fig_lc}
\end{figure*}

Explicit in our calculations is the assumption that we can solve
for the level populations and temperature structure
neglecting the relativistic and the time dependent terms
in the radiative transfer equation.
While it is moderately easy to include the relativistic terms
(e.g., Hauschildt 1992) the time dependent term is equally important
and must be included for consistency \citep{Pinto_Eastman_2000a,Pinto_Eastman_2000b}.
In hydrodynamic models, both terms are routinely
(and must be) included, both for the hydrodynamics and computing the light curve.

However for spectral calculations it is usual to follow
a stellar approach and assume that we
can compute a static atmosphere by specifying the
luminosity via the diffusion approximation at the lower
boundary. Studies of the importance of the relativistic terms have
been done, for example, by Hauschildt et al. (1991), or by Jeffery (1993),
and shown to be important. What is not clear, however, is what effect
the terms will have on the distances based on the Expanding Photosphere Method
(Dessart \& Hillier 2005b), something that we are currently investigating.
Very preliminary work suggests that the inclusion of relativistic terms,
while producing spectra that are very similar, could lead to changes in EPM
distance by up to 10\%. However, we do not expect that the inclusion of these
terms will effect the conclusions in this paper; the photospheric temperatures
are similar and the importance of the time-dependent terms arises primarily through atomic physics,
and should not be sensitive to radiative transfer affects.

   \subsection{Presentation of models}
\label{sect_models}

\begin{figure*}
\epsfig{file=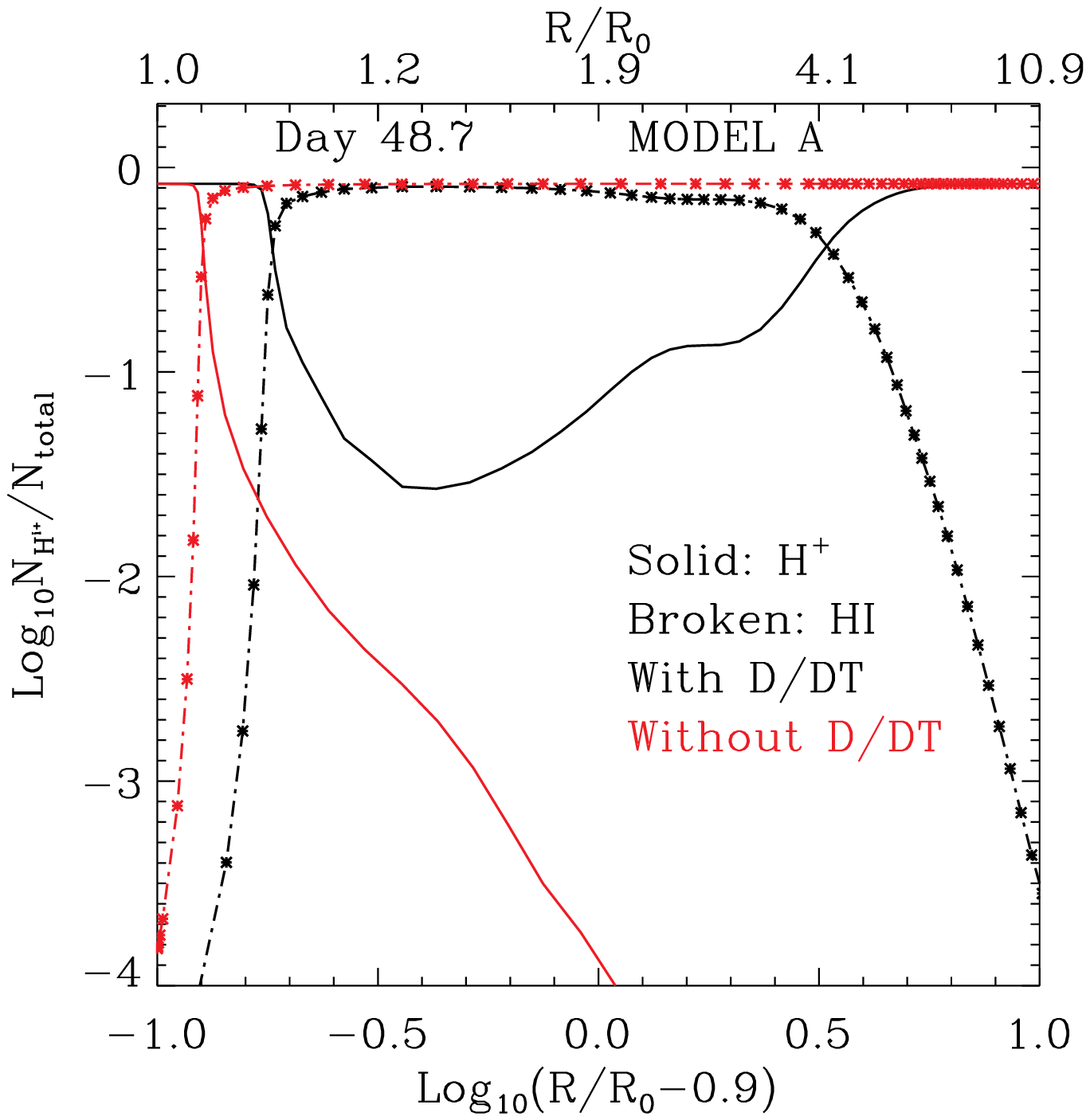,width=8cm}
\epsfig{file=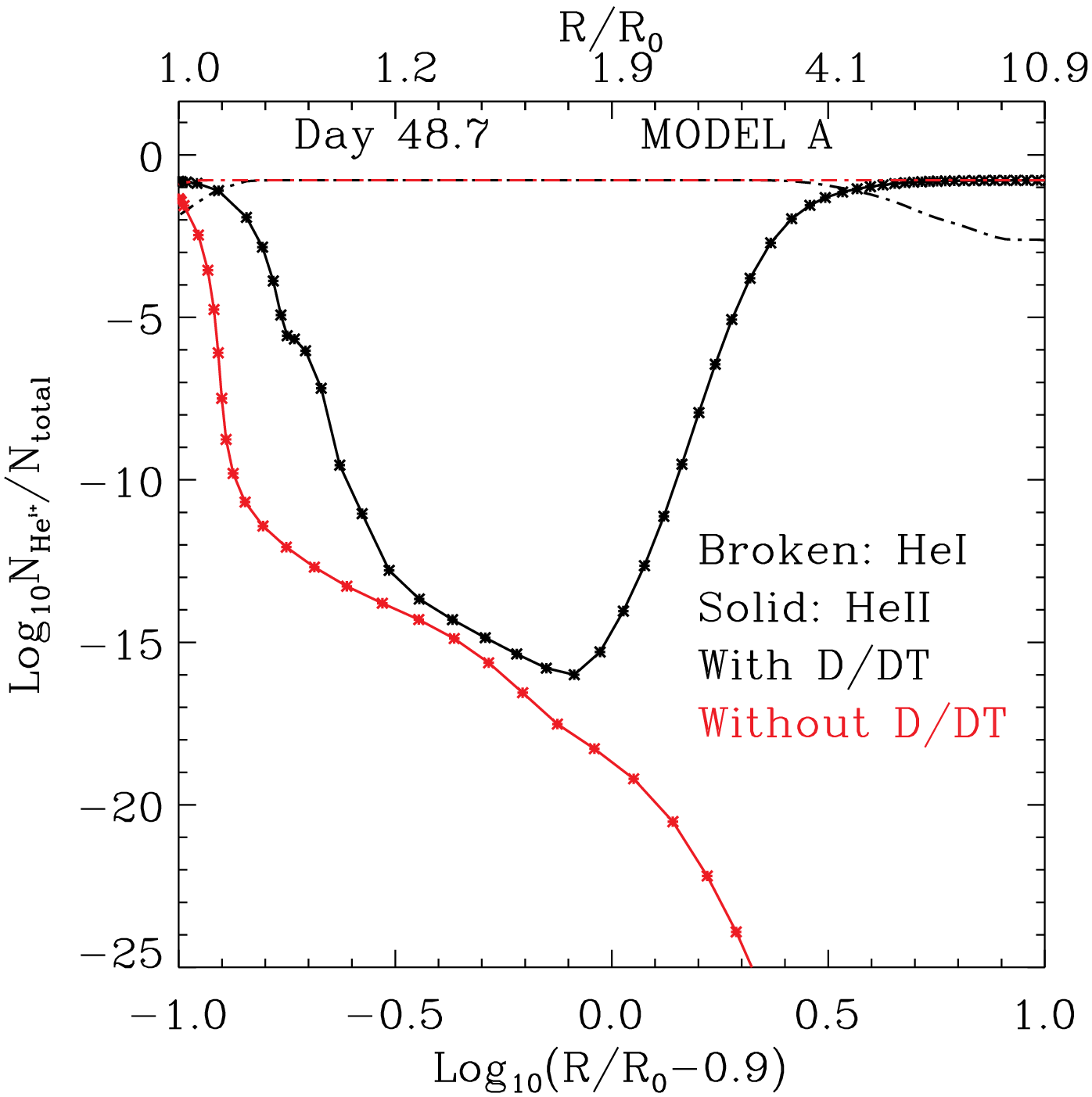,width=8cm}
\caption{
{\it Left}: Radial variation of the ionization fraction of Hydrogen
(H{\sc ii}: solid line; H{\sc i}: Broken line) for the baseline model A with
time dependence (black) and without (red), and at 48.7 days after explosion,
but otherwise having {\it identical} parameters.
Note the over-ionization of hydrogen at large radii, not present
in its neutral state despite the very late time, akin to a frozen-in ionization.
Symbols give the positions of the CMFGEN adaptive-grid points, for each model but for only
one curve.
Note the presence of the well-resolved recombination front near the base,
deeper-in for the steady-state CMFGEN model.
{\it Right}: Same as left, but for the ionization fraction of helium
(He{\sc ii}: solid line; He{\sc i}: Broken line). Notice the same frozen-in
ionization at large radii.
[See the electronic edition of the Journal for a color version of this figure,
and the text for discussion.]
}
\label{fig_h_he}
\end{figure*}

    We adopt a different procedure to model SN spectra from that employed previously
in \citet{DH_05a, DH_05b, DH_06a} by modeling the time evolution of a larger
portion of the ejecta, and in a Lagrangean spirit.
Each model is stepped in time using
the implicit method outlined above. We have tested the dependence of
our results on the adopted time increment.
Two sequences evolved with a 1 day increment or a 2 day
increment, but starting and ending with the same model input
parameters, yield an SED at the final time that agree to better than 0.1\% on average,
departures peaking at 1\% at certain UV wavelengths (coinciding with strong lines
but associated with spectral regions of negligible flux).
In practice, we increase the time increment (slightly) with the time elapsed since explosion,
using a fraction of a day within the first week, but $\sim$3 days
at a few weeks after shock breakout. For simplicity,
we choose the same minimum and maximum
velocities for all models in a given sequence.
Because we neglect the relativistic and time dependent terms
from the transfer equation, the base luminosity is a model parameter
only, and does not reflect the true physical base luminosity.
Implicit in our method is the assumption that the observed
spectrum primarily depends on the local photospheric conditions
as set by the observed luminosity. We use the same approach in
the steady-state modeling, although in that case we only need to
model the atmosphere to a depth where we can utilize the
diffusion approximation, usually corresponding to a Rosseland
optical depth of a few tens. Here, the initial inner radius
of the computation is chosen so that it remains optically thick in the continuum over
the entire time sequence, thereby ensuring at all times the adequacy of the diffusion
approximation employed at this inner boundary. (Note that we refer to the photosphere
as the location where the inward-integrated continuum optical depth equals two thirds.)
To isolate the effects associated with time dependence, we neglect any chemical
stratification (homogeneous ejecta) and assume a uniform power-law density
distribution ($d \ln \rho / d \ln R = -n$), with the exponent $n$ set to 10.

   We illustrate time-dependence effects with four different models.
We set up a ``baseline'' model (A) that mimicks the properties of SN 1999em \citep{DH_06a}.
In practice, we initially map onto the CMFGEN grid an envelope with the following properties:
$R_0 = 2.524 \times 10^{14}$\,cm ($R_0$ is the base radius), $v_0 = 3470$\,\kms
($v_0$ is the base velocity), $R_0^3 \rho_0 = 6.781 \times 10^{33}$\,g
($\rho_0$ is the base density), $\rho(R) = \rho_0 \left( R_0/R \right)^n$ with the
density exponent $n=10$ ($R$ is the radius), and $L_0 = 3.6 \times 10^{11}$\lsun ($L_0$
is the base luminosity). The maximum radius is 16 times the value of the base radius $R_0$
(this choice ensures that
at the outer CMFGEN boundary, the radiation field is more decoupled from matter at all frequencies
and free streaming, i.e., the Eddington factor is close to unity).
These initial conditions correspond to a base Rosseland optical depth of 3640, a total
mass of 6.17\,\msun, and, assuming homology, a time after explosion of 8.4 days.
We adopt a fixed composition (i.e,. chemical homogeneity), with values consistent with
the red-sugergiant progenitor of SN 1999em \citep{DH_06a, Smartt_etal_2002}.
Adopted abundances, given by number, are: H/He = 5, C/He = 0.00017, N/He = 0.0068, O/He = 0.0001,
and all metal species are at the solar value.
We also set up a slow (B; SN 1999br-like) and a fast (C) model, differing in the velocity of its inner mass shell,
i.e., 2000\,\kms (4400\,\kms) rather than 3471\,\kms. The inner radius is also adjusted
so that we start in these three cases at about one week after explosion.
Finally, we set up an ``early-time'' model (D) that starts
at $\sim$2\,days after explosion. In this model, we employ a slightly different composition
for CNO elements, with C/He = 0.0001, N/He = 0.0013, O/He = 0.0016 (Dessart et al. 2007).
We give a synopsis of all these model characteristics in Table~\ref{table_model_param}.
Apart from the ``early'' model that covers only the first week, we evolve
these models from one week up to about 6 weeks after shock breakout.

\begin{figure*}
\epsfig{file=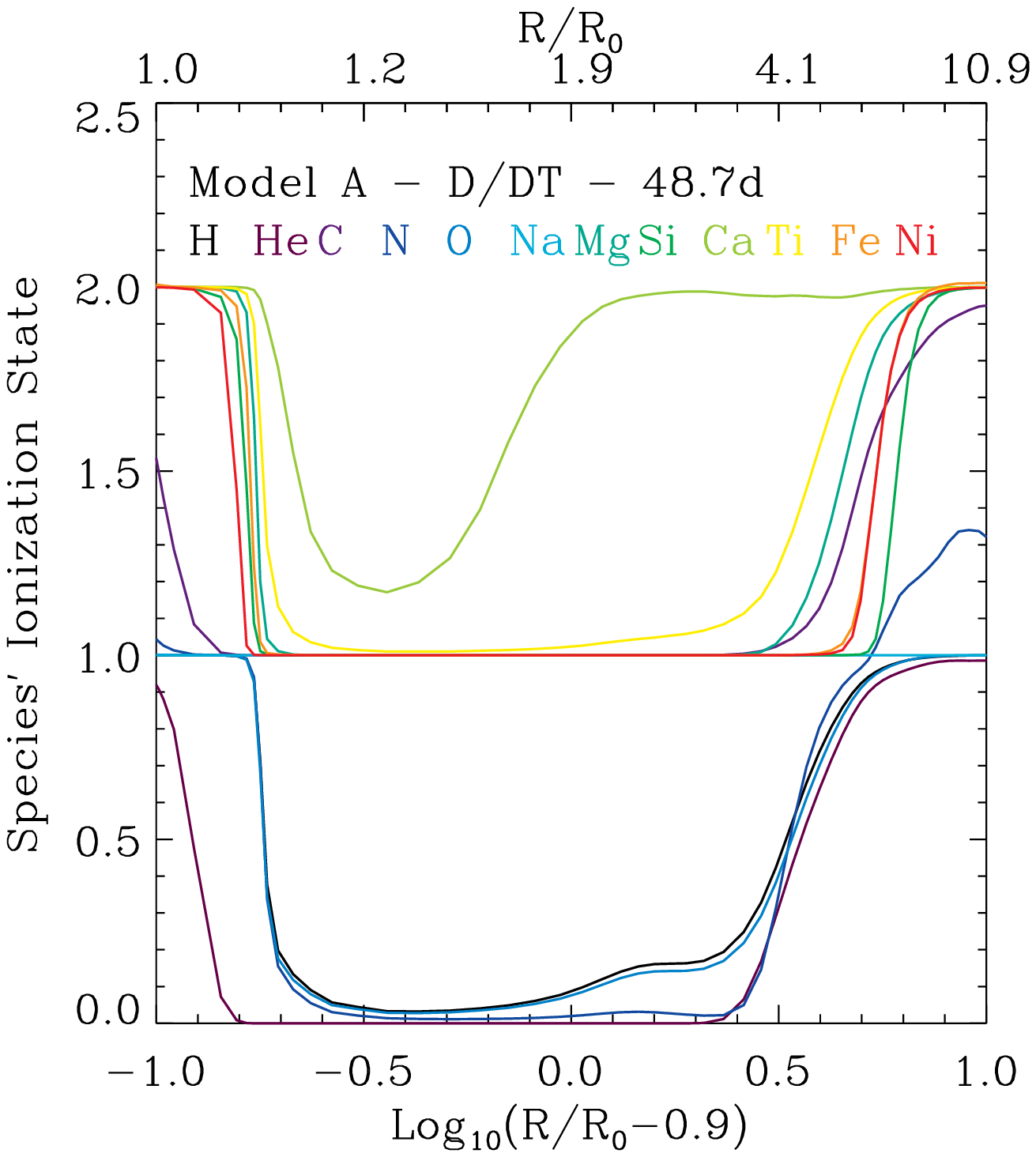,width=8cm}
\epsfig{file=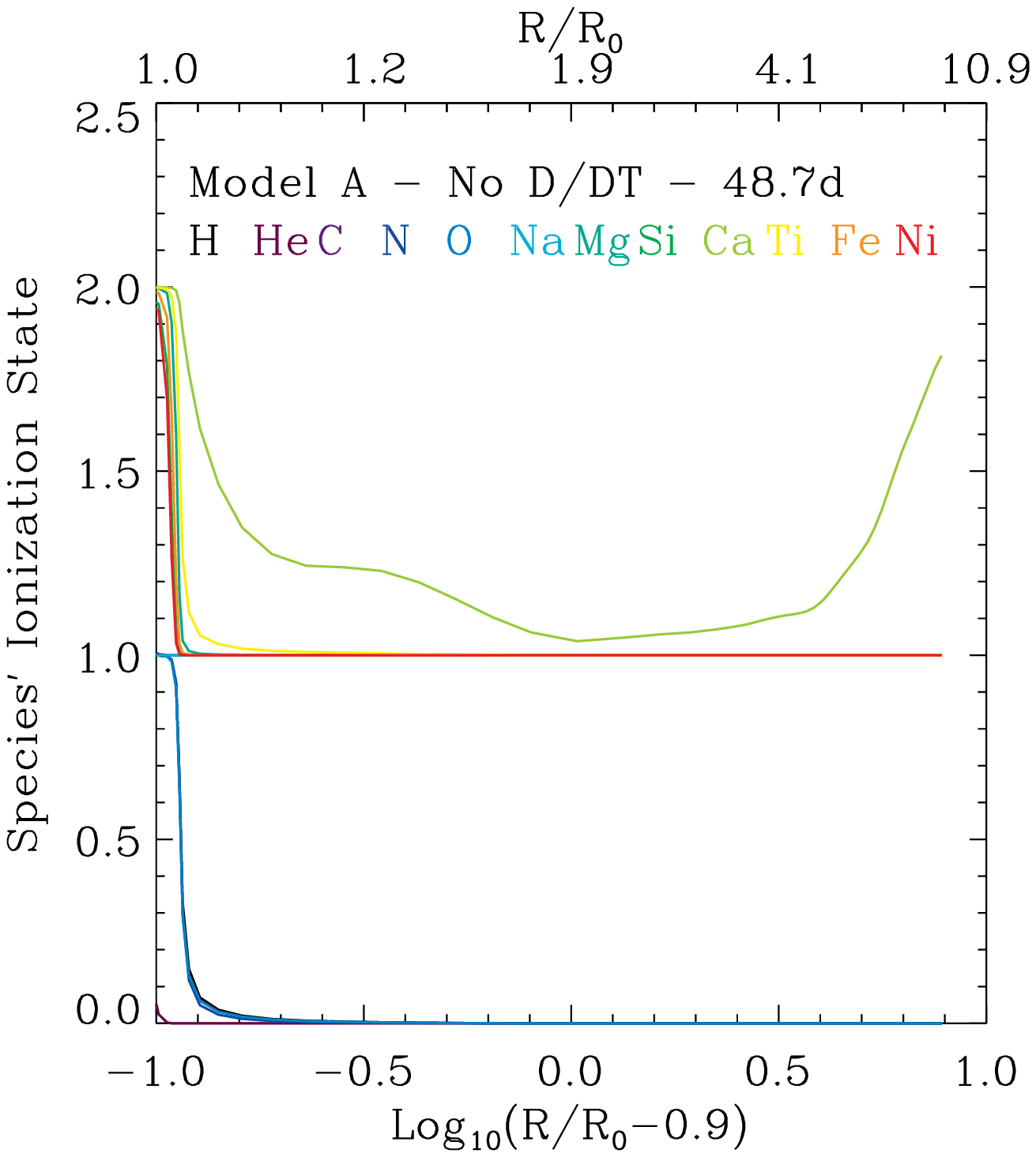,width=8cm}
\epsfig{file=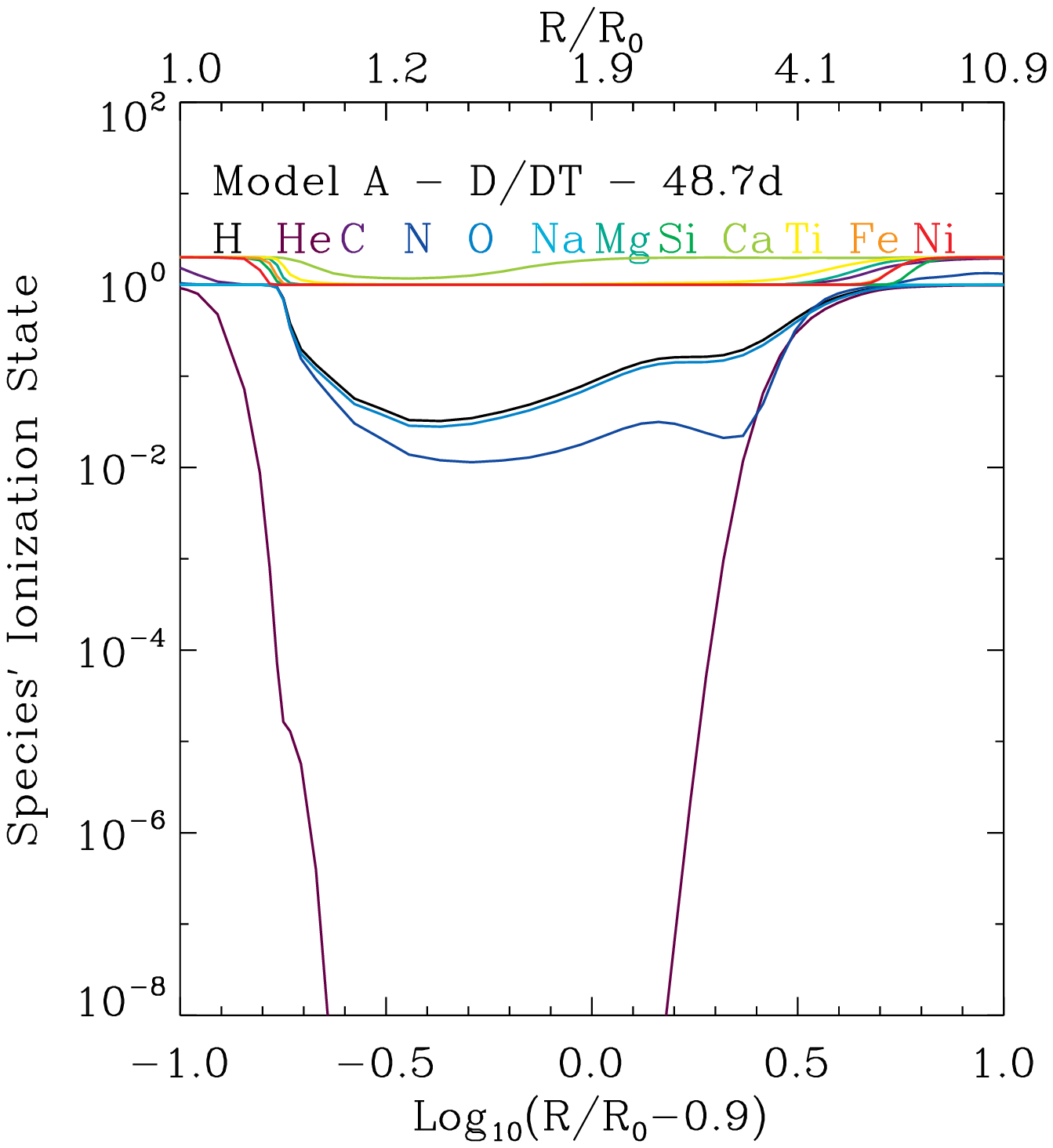,width=8cm}
\epsfig{file=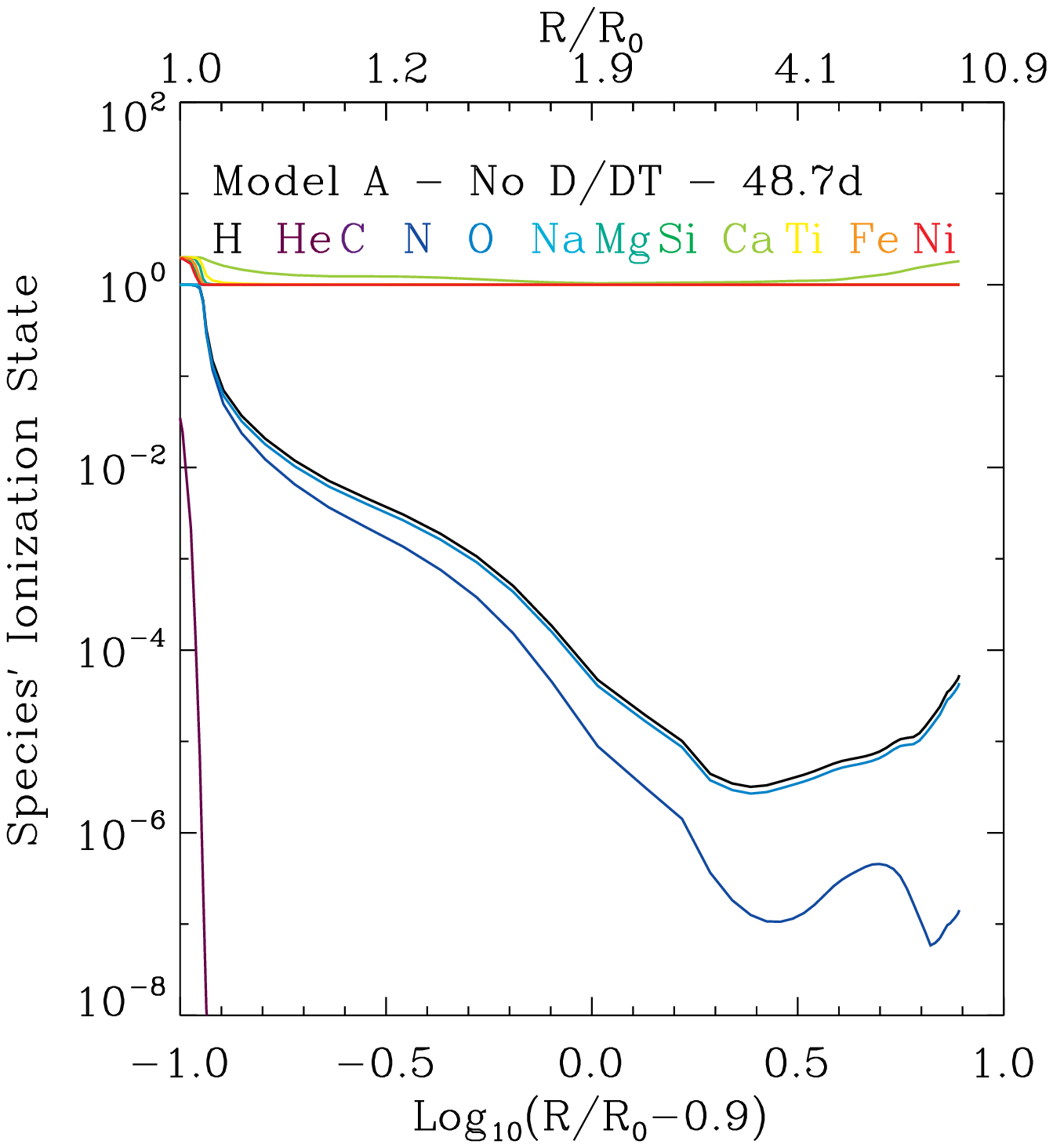,width=8cm}
\caption{
{\it Top row:} Comparison of the radial variation of the ionization state
of all species included in model A, for the time-dependent (left) and the
steady-state case (right). Note how the effects on hydrogen and helium
discussed in \S\ref{effects_on_ejecta} and illustrated in Fig.~\ref{fig_h_he}
apply in fact to all species, with a systematic over-ionization at large
distances.
{\it Bottom row:} Same as for the top-row panels, but this time shown in the
log to reveal better the magnitude of the predicted difference in ejecta ionization
between time-dependent and steady-state approaches.
[See the electronic edition of the Journal for a color version of this figure,
and the text for discussion.]
}
\label{fig_ion_state}
\end{figure*}

In our current approach, each ion-level population is evolved in a time-dependent non-LTE fashion,
and thus, for simplicity, we adopt a unique model atom for all steps in a given model
time sequence. For the baseline, slow, and fast models, we treat
H{\,\sc i} (30,20), He{\,\sc i} (51,40), He{\,\sc ii} (5,5),
C{\,\sc ii} (59,32), C{\,\sc iii} (20,12), C{\,\sc iv} (14,9), N{\,\sc i} (104,44),
N{\,\sc ii} (41,23), N{\,\sc iii} (8,8), O{\,\sc i} (75,23), O{\,\sc ii} (111,30),
O{\,\sc iii} (46,26), Na{\,\sc i} (71,22), Mg{\,\sc ii} (65,22), Si{\,\sc ii} (59,31),
Si{\,\sc iii} (53,29), Ca{\,\sc ii} (77,21), Ti{\,\sc ii} (152,37),
Fe{\,\sc ii} (309,116), Fe{\,\sc iii} (477,61), Fe{\,\sc iv} (282,50), and Ni{\,\sc ii} (93,19),
where the parenthesis contains the number of full- and super-levels
(see Hillier \& Miller 1998 for details).
Compared to the model atom used in \citet{DH_06a}, we omit C{\sc i}, neon, aluminium, sulfur,
chromium, manganese,  and cobalt.
Given the much higher ionization state at earlier epochs, the ``early'' model neglects most
low-ionization species (and metals other than iron) and adds He{\,\sc ii} (30,13), C{\,\sc iv} (64,59),
N{\,\sc iv} (60,34), N{\,\sc v} (67,45), O{\,\sc iv} (72,53), O{\,\sc v} (152,75),
Fe{\,\sc v} (191,47), and Fe{\,\sc vi} (433,44).

From a converged initial steady-state model, we then step in time by
$\Delta t = 3$\,days (0.7\,day for the early model D), i.e., all mass shells $m$ in the initial model
are evolved from their position $R(m)$ to a new radius $R'(m)$, with $R'(m) = R(m) + v_0(m) \Delta t$,
a new density $\rho' = \rho (R/R')^3$, and the same velocity.
In practice, for each new model, CMFGEN adapts its 80-point spatial grid to cover
each optical-depth decade with at least five points. For model D, the grid contains
only 50 radial points.

In \S\ref{comp_with_obs}, we present a sample of comparisons to observations to {\it illustrate}
the performance of these time-dependent models (note that our models were not tailored to match
any specific observation). For the ``early'' model D, observations are only starting now to be available through, e.g.,
the SWIFT satellite (Brown et al. 2007), and the corresponding simulations for model D, thus,
remain largely unconstrained (but see Quimby et al. 2007; Dessart et al. 2007).
To illustrate the bolometric luminosity and color evolution of our four models, we
present in Fig.~\ref{fig_lc} their light curves covering from the UV to the near-IR,
with properties quite typical of Type II-P SNe (Leonard et al. 2002;
Pastorello et al. 2006; Dessart et al. 2007).


\section{Results}

   \subsection{Time-dependence effects on ejecta properties}
   \label{effects_on_ejecta}

\begin{figure}
\epsfig{file=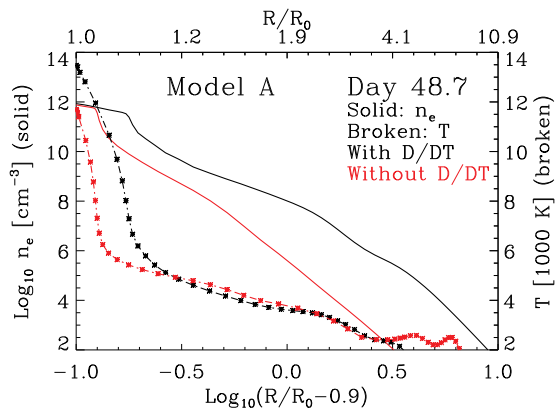,width=8cm}
\caption{
Radial variation of the electron density $n_e$ (solid line; in the logarithm, and in units of cm$^{-3}$)
and the temperature (broken line; in units of 1000\,K), for the time-dependent baseline
model (black), and its steady-state equivalent (red).
Both models have otherwise identical parameters and correspond to a time of 48.7
days after explosion.
[See the electronic edition of the Journal for a color version of this figure,
and the text for discussion.]
}
\label{fig_t_ed}
\end{figure}

  We start our discussion of time-dependence effects on Type II SNe by focusing
on the properties of the ejecta. In Fig.~\ref{fig_h_he}, we show the ionization
fraction for hydrogen (left) and helium (right) for model A at the end
of the time sequence, 48.7\,days after the explosion.
Hydrogen recombination occurs just above the photosphere but
less efficiently in the time-dependent model than in the steady-state model,
neutral hydrogen being 3 orders of magnitude more abundant in the later model.
In the time-dependent model, two additional features of interest are the presence of
fully ionized hydrogen at large distances above the photosphere (the hydrogen ionization
state is frozen) and the delayed recombination in the optically thick layers,
at and below the photosphere (the ionization front is located further out).
All these features apply to the helium ionization fraction as well, which we show in
the right panel of Fig.~\ref{fig_h_he}. The He{\sc ii} population is 15 orders
of magnitude lower than that of He{\sc i} at intermediate heights, but it is comparable
or stronger at large distances. Effects of time dependence can thus reach
considerable levels, much beyond modulations that can be achieved through reasonable changes of
(or uncertainties in) chemical abundance or ionizing fluxes.


Because metal line-blanketing plays an
increasingly important role at and after the recombination epoch, time-dependence effects on metals
are also of special interest.
We show in Fig.~\ref{fig_ion_state} the ionization state of all species for model A on
day 48.7. Notice how the frozen-in ionization is readily visible at large distances above the
photosphere, with order of magnitude differences in ionization level between the time-dependent
and steady-state approaches. Hence, the inclusion of time-dependence
affects all species, irrespective, for example, of the atomic properties of
the given ion, and with a magnitude that is greatest at large distances above the photosphere where
the recombination timescale becomes increasingly large compared to the expansion timescale.


  Finally, we show  in Fig.~\ref{fig_t_ed} the radial variation of the temperature and
the electron density for model A at 48.7 days after explosion.
The temperature in the time-dependent model is higher in the optically-thick
regions, due to the reduction of density-sensitive line emission processes that dominate
radiative cooling. In the optically-thin regions, above the photosphere,
the temperature is, however, quite similar
between the two approaches. This result stems from the poor coupling between gas and
radiation in these scattering-dominated, tenuous, ejecta. Departures from LTE are pronounced and the
temperature is a secondary driver of the ejecta ionization, even in steady-state
configurations.

  Models B-C-D offer valuable information on the sensitivity of the above effects on
the ejecta velocity (models B and C) and the epoch (model D). We find that
varying the velocity of the ejecta within the range of values for slow/underluminous (model B; SN 1999br) and
more standard luminosity Type II SN events (SN 1999em) has no sizeable effect on the main features
of time dependence highlighted above. Recombination is inhibited, and a significant over-ionization
persists far above the photosphere.

\begin{figure*}
\epsfig{file=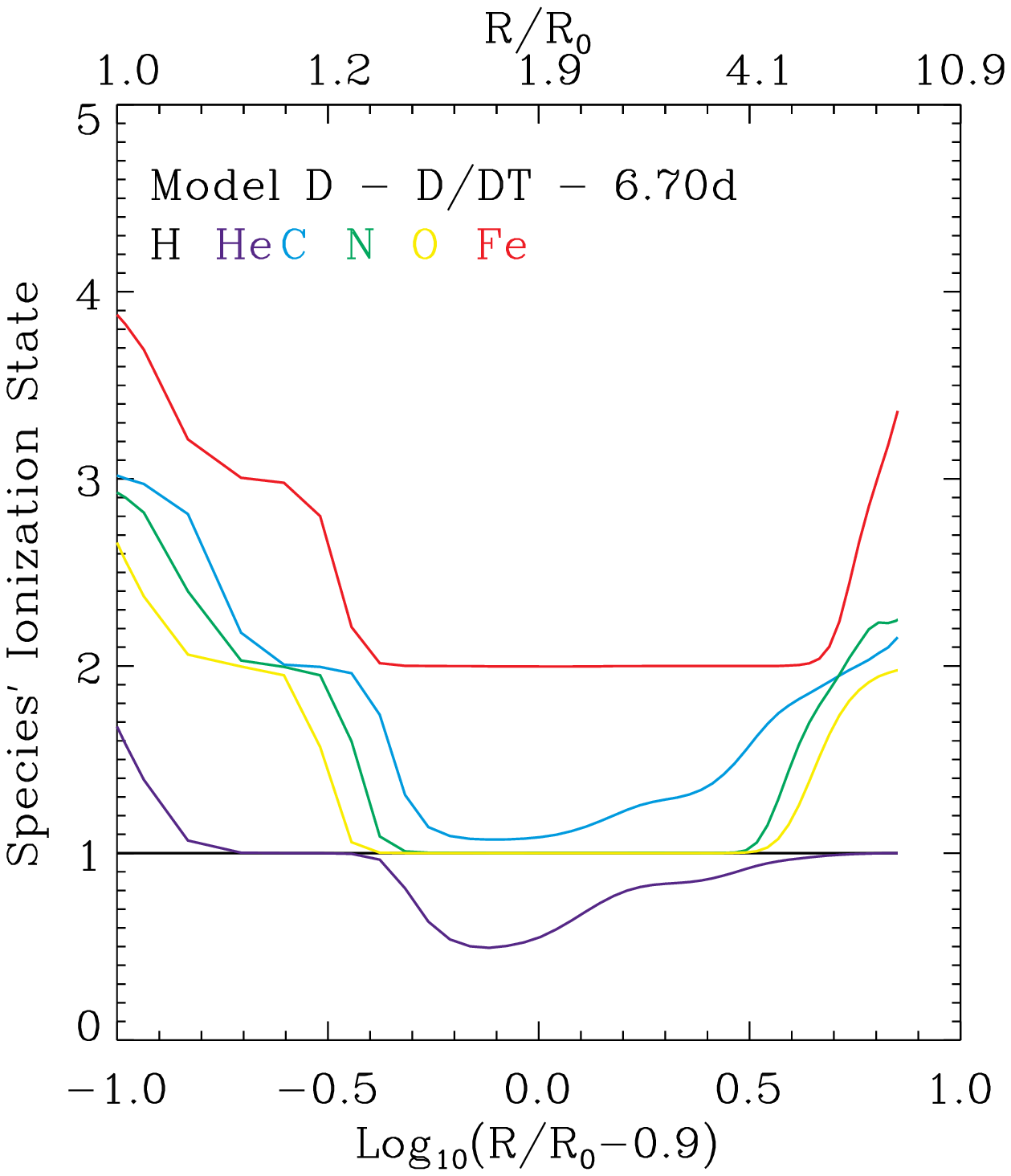,width=8cm}
\epsfig{file=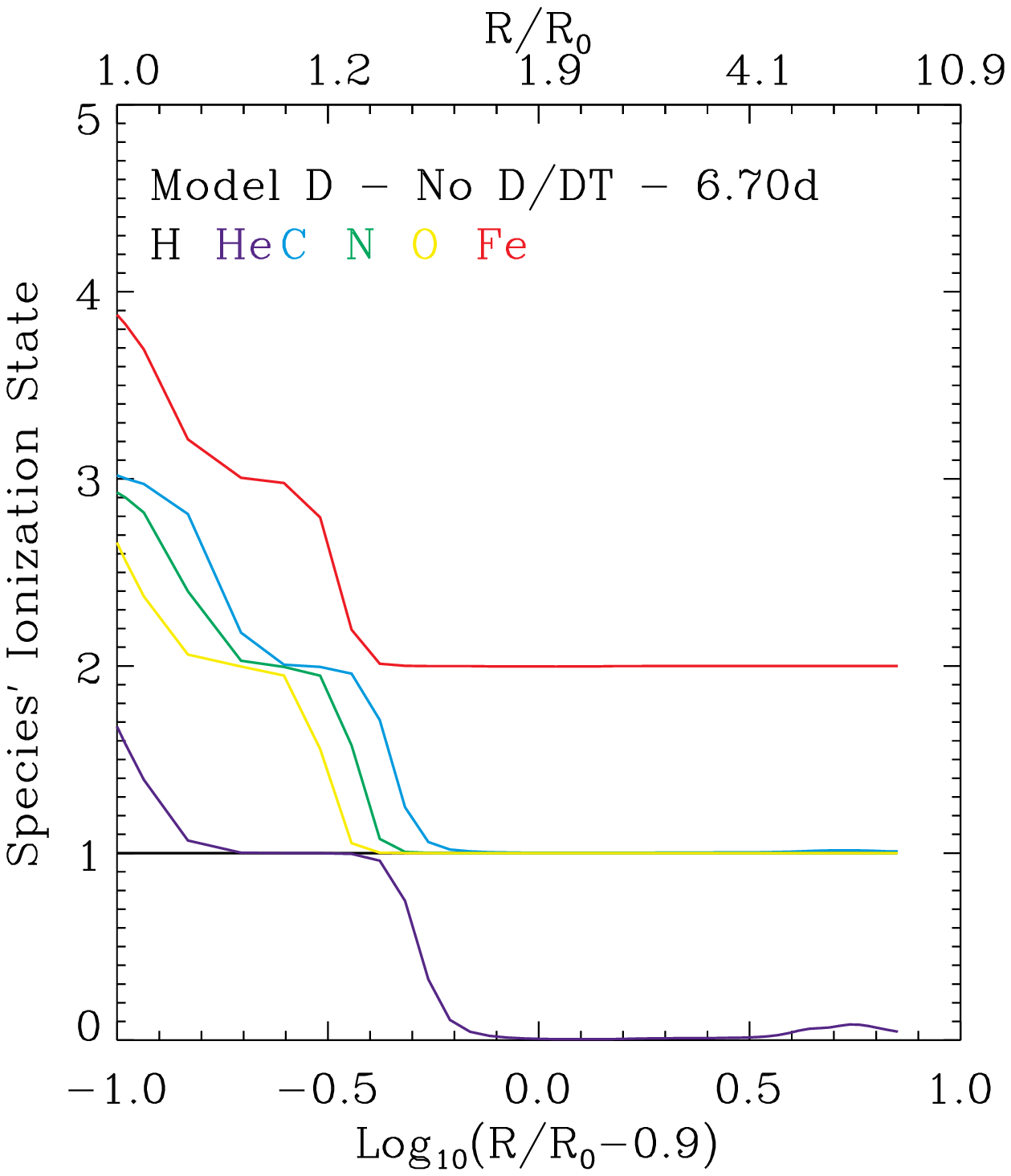,width=8cm}
\caption{
Comparison of the radial variation of the ionization state
of all species included in model D (the model that covers the first week of evolution
of a Type II SN ejecta), for the time-dependent (left) and the
steady-state case (right). Note how the behaviour illustrated in Fig.~\ref{fig_h_he}
for hydrogen and helium is visible in fact in all species (with the exception of hydrogen
which remains fully ionized in both time-dependent and steady-state approaches at such
an early time; black line).
Hence, even during the first week after explosion, when the ejecta density are
relatively higher, time-dependence effects leave an obvious imprint on the
ejecta ionization state (He, C, N, O, Fe here), even though the medium,
made-up primarily of fully-ionized hydrogen, retains roughly the same electron density.
[See the electronic edition of the Journal for a color version of this figure,
and the text for discussion.]
}
\label{fig_ion_state_early}
\end{figure*}

In Fig.~\ref{fig_ion_state_early}, we present the
mean ionic charge  for the six species (H, He, C, N, O, and Fe) included in model D (early epoch),
and at 6.7 days after explosion.
While hydrogen remains fully ionized, in both time-dependent and steady-state approaches, all other
species show a higher ionization at large distances above the photosphere; no difference
is noticeable in the vicinity and below the photosphere, contrary to model A at 48.7 days.
This supports the idea  
that time-dependence effects operate even at
early times. They do so at large distances because there, the recombination and expansion timescales
are already sufficiently close, but they can also act at the photosphere because
of optical depth effects and meta-stable levels (see \S\ref{sect_ddt_tau}).
It also shows that over-ionization at large distances
does not stem by essence from the recombination energy (since most of that energy is stored in
hydrogen and hydrogen remains fully ionized in this whole sequence) but results exclusively from
the low-density fast-expansion of the SN ejecta.
A corollary is that this progressive ``over-ionization'' does not require
an extra source of ionization and excitation from, for example, $^{56}$Ni (Mitchell et al. 2001;
Baron et al. 2003), and operates instead, in a robust fashion, through the reduction of
the effective recombination rates. Finally, this supports the idea that
similar effects should apply in SN ejecta {\it in general}, since they generically
possess similar low density and large velocities.

\begin{figure}
\epsfig{file=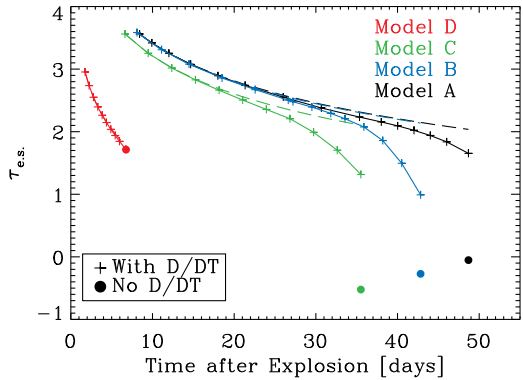,width=8cm}
\caption{
Time evolution of the electron-scattering optical depth $\tau_{\rm e.s.}$
for each time-dependent model A (black; baseline model), B (blue; slow model), C (green; fast model),
and D (red; early epoch model), using solid lines and cross symbols (to mark the days computed).
We overplot the electron scattering optical depth for the corresponding {\it steady-state}
model at the final time in each sequence (filled circle).
For each model, we also draw the $1/t^2$ optical-depth evolution for an ejecta
with fixed electron-scattering opacity (fixed composition) and ionization (dashed line;
the blue and black dashed curves overlap, and so do the red-dashed red-solid curves).
[See the electronic edition of the Journal for a color version of this figure,
and the text for discussion.]
}
\label{fig_tau_es}
\end{figure}

We conclude this section by presenting the evolution of the electron-scattering
optical depth at the base radius $R_0$ of each model in Fig.~\ref{fig_tau_es}
(see Table~\ref{table_model_param} for model parameters). Rather than having
a $1/t^2$ dependence (dashed lines) for an expanding ejecta with a fixed mass absorption
coefficient (as in model D where the dashed and solid lines overlap;
fixed composition and ionization), the curves for models A, B, and C show a
change of concavity, with a slow decrease at first and a turn-over to a faster decrease
after 3--6 weeks. Recombination, which is delayed at first and permits the persistence
of a high free-electron density in the ejecta, does in the longer term occur and triggers
a fast drop of the base electron-scattering optical depth. It remains, however, higher
by up to a factor of about ten, than the value obtained in the corresponding steady-state
model (for the same adopted lower bound radius in the optical-depth integral) at the ultimate
time in the sequence, shown as filled colored circles.
Model D, whose ejecta contains species at least once ionized throughout the sequence,
has a base optical depth of 51.9 in the time-dependent approach and 51.7 in the steady-state
approach, the modest difference stemming from the slight over-ionization of sub-dominant species
like helium and CNO elements.
(See the last columns of Table~\ref{table_model_param} for a log of these values.)

   \subsection{Time-dependence effects on the emergent light, from UV to near-IR,
for the baseline model A}
   \label{effects_on_light}

   The various investigations and analyses discussed above indicate that time-dependence
effects are robust, apply to all species, even prior to the recombination epoch of hydrogen.
Hence, for example, during the first week after explosion, such effects inhibit the
recombination from Fe$^{4+}$ to Fe$^{3+}$ and Fe$^{2+}$, or that of He$^{+}$ to He$^{0+}$.
Different species block light at specific wavelengths, with often a tendency to bunch-up
in specific spectral regions, but with no {\it a-priori} correlation, and therefore
we anticipate that such modulations in the ionization state of the ejecta might lead to a significant
impact on the properties of the emergent light. Both the optical, dominated by H{\sc i} and He{\sc i}
lines, and the UV, dominated by metal line-blanketing due to, for example, iron, should be affected.

\begin{figure*}
\epsfig{file=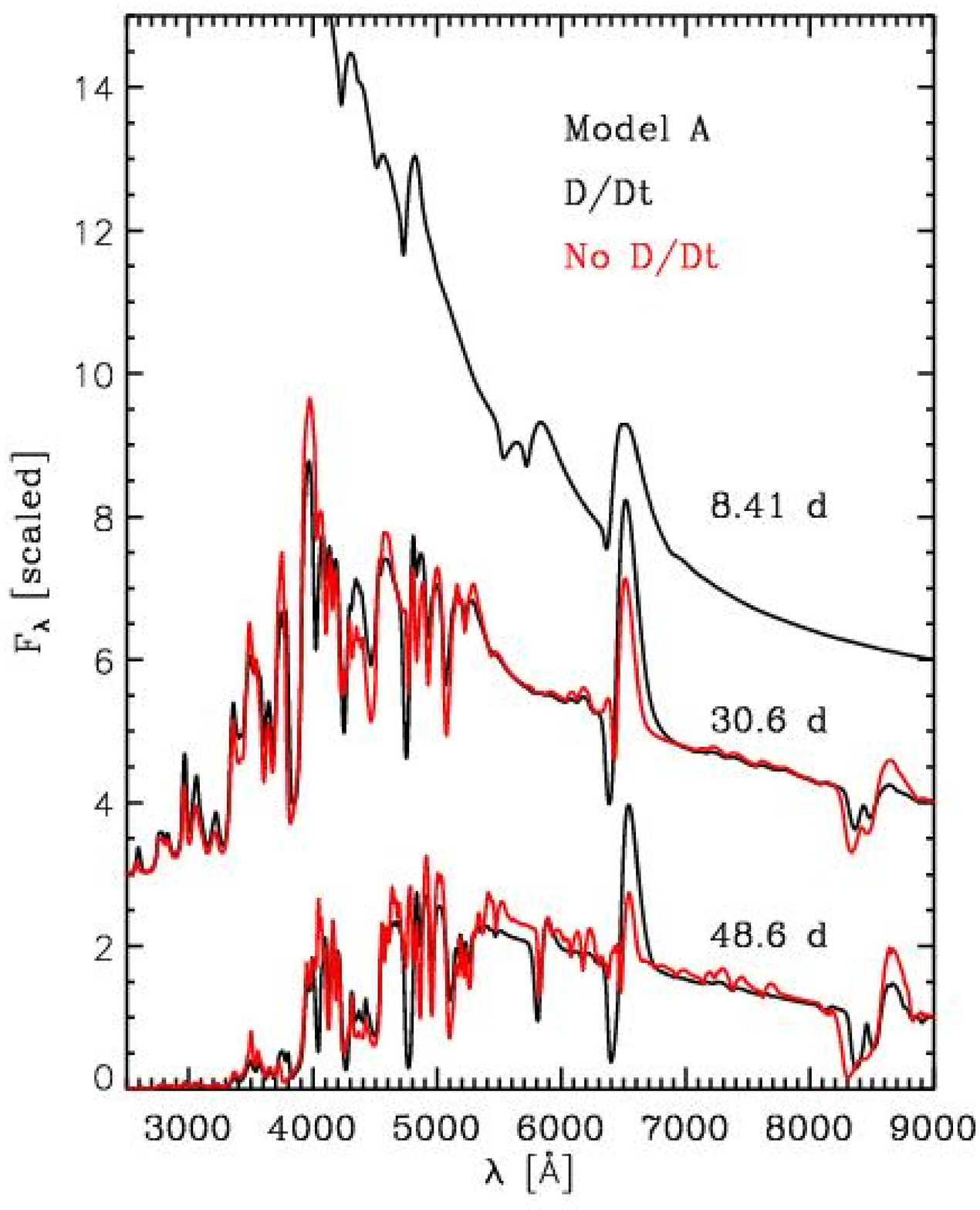,width=5.5cm}
\epsfig{file=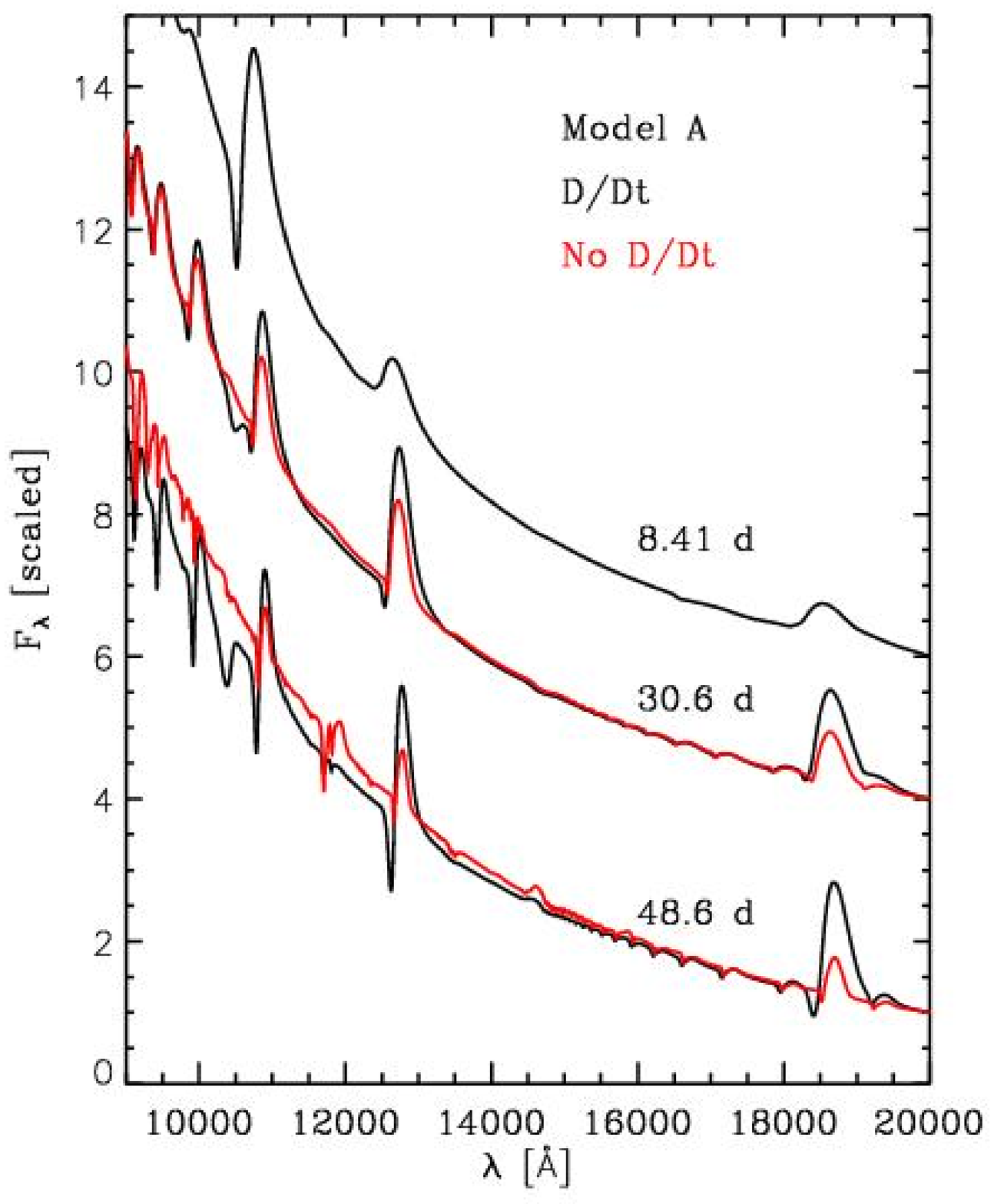,width=5.5cm}
\epsfig{file=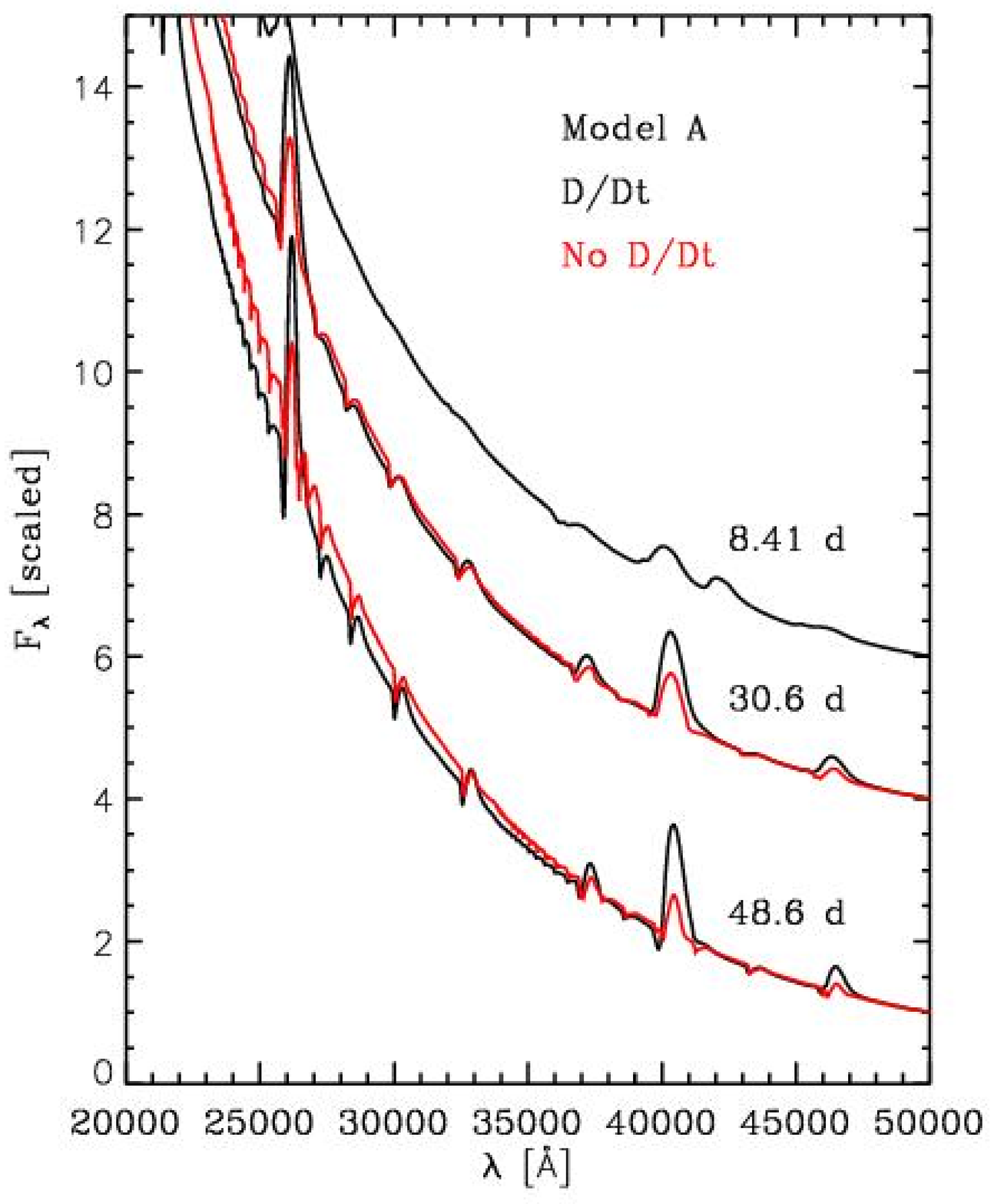,width=5.5cm}
\caption{
Montage of (scaled) synthetic spectra showing the time evolution for the time-dependent (black)
and steady-state (red) CMFGEN model A in the UV/optical (left panel), the 1--2$\mu$m region (middle panel),
and the 2--5$\mu$m region (right panel). A vertical shift is applied to each spectrum,
smaller with increasing time. We show the evolution for three dates,
at 8.41\,d (top curve; the time-dependent and the steady-state
both start their evolution at that time and, thus, only one curve is plotted),
slightly past halfway through the time sequence at 30.6\,d (middle curve), and for the last
time computed at 48.6\,d (bottom).
[See the electronic edition of the Journal for a color version of this figure,
and the text for discussion.]
}
\label{fig_spec_modelA}
\end{figure*}

  In Fig.~\ref{fig_spec_modelA}, we present a montage of synthetic spectra
showing the time evolution for the time-dependent (black)
and steady-state (red) CMFGEN model A in the UV/optical (left panel), the
9000\,\AA\--2$\mu$m region (middle panel),
and the 2--5$\mu$m region (right panel) at three consecutive
epochs (8.41\,d, the start of the sequence, 30.6\,d, and the last time in the sequence at 48.6\,d).
The SED has a peak that shifts from the UV to the optical
over the 6 weeks covered, starting first as a nearly featureless continuum and
then revealing more and more the effects of metal line blanketing in the UV and optical
(see, e.g., Brown et al. 2007). At a {\it qualitative level}, this model sequence, which accounts
for time dependence, is in agreement with what is observed in nature, for example for
the Type II-P SN 2006bp (Dessart et al. 2007), and what could also be modeled with steady-state
CMFGEN models.
Hence, the time-dependence incorporated here has no major influence on the general
morphology of the SED and its evolution with time.
A corollary is that time-dependence effects on the continuum SED are weak.
The effects on lines are, however, pronounced as we notice that all hydrogen lines,
from the Balmer to the Paschen and
to the Brackett series, are considerably weaker in the steady-state model for the last
date, compared to its time-dependent equivalent, suggesting that effects do
not intervene selectively, but rather that all hydrogen lines are affected.

\begin{figure}
\epsfig{file=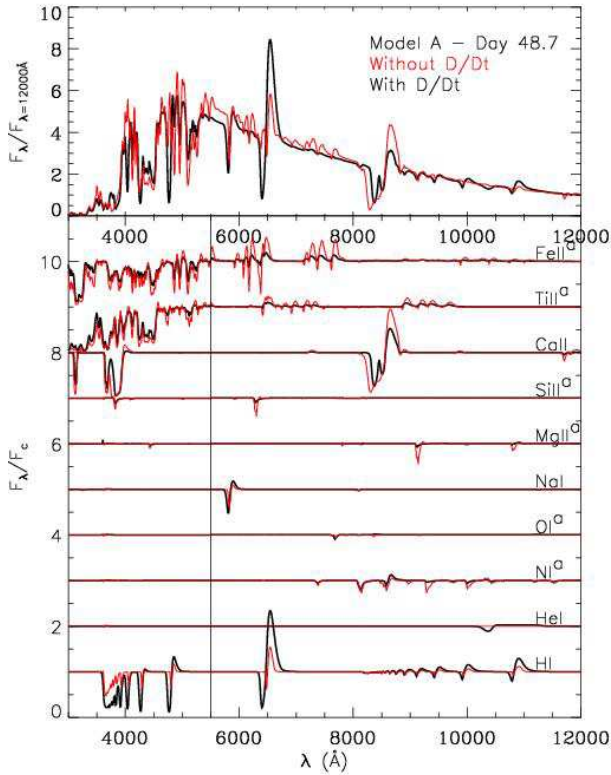,width=8cm}
\caption{
{\it Top:} Comparison of synthetic spectra obtained for model A and the last time in the
sequence at 48.7 days, showing the results for the time-dependent model (black curve)
and the equivalent steady-state model on the same day (red curve).
{\it Bottom:} Same as top, but for the synthetic rectified spectra (shifted vertically
for visibility, and magnified by a factor of five beyond 5500\AA\ for all species
labelled with a superscript $a$)
and allowing exclusively for bound-bound transitions of the ion species indicated on the
right. We only show those species that leave an imprint on the spectrum.
This figure illustrates the modest time-dependence effects over the first 6 weeks
after explosion on the continuum. However, the effects on lines are significant,
most noticeable on the strengthened hydrogen Balmer lines and Na{\sc i}\,D, and the weakened
Ca{\sc ii}\,8500\AA\ multiplet lines.
The enhanced recombination of the ejecta in the steady-state model results in
a lower ionization state and stronger metal lines, e.g., for Fe{\sc ii} and Ti{\sc ii}.
[See the electronic edition of the Journal for a color version of this figure,
and the text for discussion.]
}
\label{fig_spec_DDT_noDDT}
\end{figure}

  To delineate more precisely the influence on lines, we display a montage of optical synthetic
spectra for model A, corresponding to the last time in the sequence, in Fig.~\ref{fig_spec_DDT_noDDT}.
In the top panel, we plot the result for
the time-dependent (black) and the equivalent steady-state (red) model, while in the lower
panel, we present rectified spectra obtained by post-processing the CMFGEN results and accounting
for all continuum processes but only bound-bound transitions of selected ions.
The contrast for Hydrogen Balmer lines is stark, and
most extreme for H$\alpha$, which is now as strong as observed.
Helium, which does not show a single line in the steady-state model, shows a non-trivial
absorption feature at $\sim$1.03$\mu$m and a very extended weak flat-topped emission profile
(we will return to this line below).
Continuing upwards, nitrogen and oxygen reveal only weak lines, and effects are marginal.
Steady-state models show stronger absorption and weaker emission in N{\sc i} lines.
The same holds for Mg{\sc ii} and Si{\sc ii}.
In the time-dependent case,
Na{\sc i}\,5895\AA\ strengthens and broadens. This occurs because
Na$^+$ is the dominant ionization state in both models, but the increased
electron density in the time-dependent model leads to an increase in the
neutral Na fraction. Conversely, Ca{\sc ii}\,8500\AA\ weakens, primarily because
the relative fractions of Ca$^+$ and Ca$^{++}$ change significantly.
Fe{\sc ii} and Ti{\sc ii} lines are stronger in the steady state model ---
stronger in absorption in the UV and stronger in emission in the optical --- resulting
from the lower ionization of the ejecta.
When combined, all these effects give an SED that
is most conspicuously different through the changes in the strength of H$\alpha$
and Ca{\sc ii}\,8500\AA, whose ratio is reversed between the two approaches.

An alternative approach of strengthening the weak H$\alpha$ profile,
through a flattening of the density distribution\footnote{Note that the recombination
efficiency scales as the product of the electron density $n_e$ with the ion density $n_i$.
Flattening the density distribution enhances the material density $\rho$ above the photosphere,
but if the ejecta is recombined, the low values of
$n_e$ and $n_i$ will inhibit the recombination needed to power, for example, a strong H$\alpha$ line
(see also \S\ref{comp_with_obs}).},
leads instead, primarily, to a
considerable strengthening of the Ca{\sc ii}\,8500\AA\ multiplet,
in complete contradiction with observations \citep{DH_05a}. Time dependence offers here a
means to produce a strong H$\alpha$ line, while delayed
recombination of calcium maintains the moderate Ca{\sc ii}\,8500\AA\ strength.
The moderate impact on metals in the vicinity of the photosphere leaves metal
line-blanketing only weakly affected, and it is only the strongest lines
(like Fe{\sc ii}\,5169\AA) that show both a strengthening and a broadening
in the less-ionized, steady-state, model.

\begin{figure}
\epsfig{file=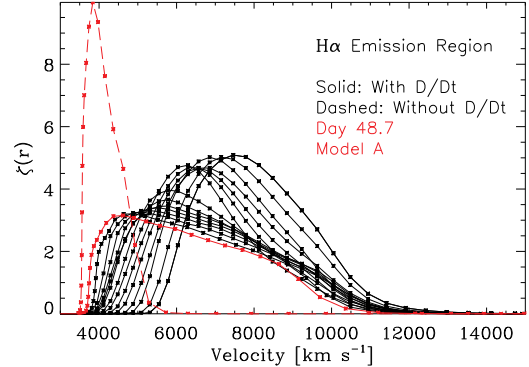,width=8cm}
\caption{{\it Solid line}: Time evolution, in velocity space and for model A, of the $\xi$
parameter of \citet{Hillier_1987} for H$\alpha$, which describes the spatial distribution of
the corresponding line emission (the area under the curve scales with the emergent line flux).
Notice how this distribution peaks at smaller radii for later times, but also the persistence
of emission at large radii at all times (the last time shown in red).
{\it Broken line}: Same as for the solid line but for the steady-state CMFGEN model,
and only for the last time (to be compared to the solid red line). Note the deeper location of the
emission, resulting in a very narrow H$\alpha$ profile, in contradiction to observations
at the corresponding mid-photospheric-phase epoch.
[See the electronic edition of the Journal for a color version of this figure,
and the text for discussion.]
}
\label{fig_ep_halpha}
\end{figure}

   In Fig.~\ref{fig_ep_halpha}, we plot the radial distribution of the parameter
$\zeta$ of \citet{Hillier_1987} for the H$\alpha$ line, a quantity that relates to
the total emission in the line through the integral
$\int_{R_0}^{R_{\rm Max}} \zeta(R') d \log R'$ (or, equivalently,
$\int_{V_0}^{V_{\rm Max}} \zeta(V') d \log V'$
for homologously expanding ejecta). The curves for the time-dependent (solid line, in
black apart from the final model in red) model A indicate
a sustained and comparable emission distribution throughout the sequence.
The peak emission shifts from 8000\,\kms at one week to 4000\,\kms at 6 weeks after explosion,
following the expansion of the ejecta
and the decrease in density. But the emission region
considerably broadens with time, so that, while its inner edge recedes, its outer edge
remains fairly fixed (in a Lagrangean sense).
The corresponding emission region for the steady-state model
(broken red line) is located deeper (the photosphere as well; see \S\ref{effects_on_ejecta})
and is considerably narrower. This property translates directly in the observed H$\alpha$
line properties, with a smaller line flux (weak line) confined to smaller velocity (narrow line)
in the steady-state model.

\begin{figure*}
\epsfig{file=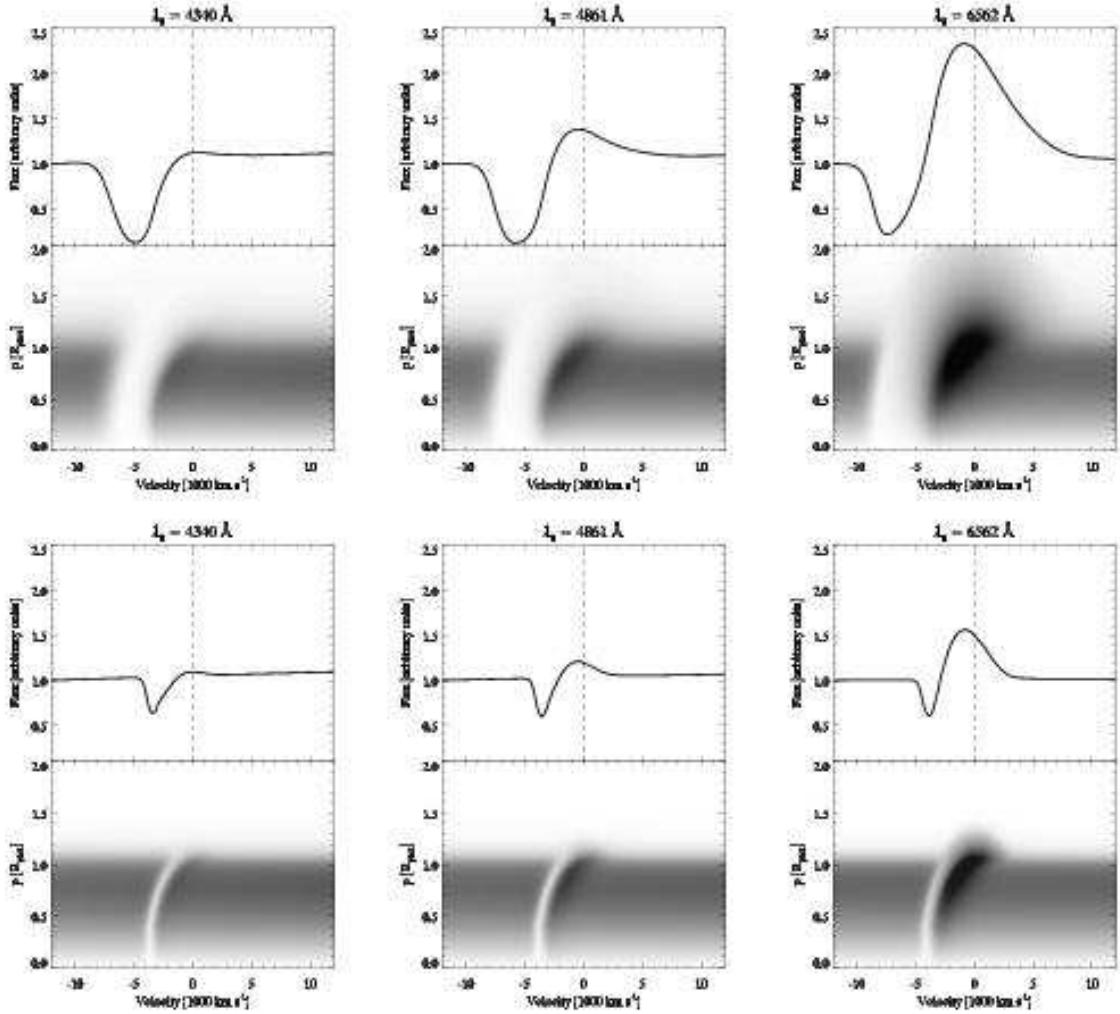,width=18cm}
\vspace{-9.cm}
\caption{
Grayscale images for H$\gamma$ (left column), H$\beta$ (middle column), and H$\alpha$ (right column),
of the quantity $p \cdot I(p)$ as a function of impact parameter $p$
and classical Doppler velocity $v=(\lambda/\lambda_0-1)c$, where $p$ is the impact parameter
(in units of the photospheric radius R$_{\rm phot}$), and $I(p)$ is the specific intensity
along $p$ (at $v$).
The panels in the top (bottom) row correspond to the time-dependent (steady-state) results
obtained for the model A ejecta at 48.7 days after shock breakout.
The top half of each panel gives the line profile flux, which, for each $v$, corresponds to
sum of $p \cdot I(p)$ over all $p$s, giving a vivid illustration of the $(p,v)$ sites
contributing to the observed line profile.
To emphasize the differences between lines and model assumptions, axis extrema are kept
identical for all frames.
}
\label{fig_balmer_pip}
\end{figure*}

To further illustrate the increase of emission-line volumes, and following \citet{DH_05a,DH_05b},
we show in Fig.~\ref{fig_balmer_pip}, for H$\gamma$ (left-column panels),
H$\beta$ (middle-column panels), and H$\alpha$ (right-column panels),
grayscale images in the $(v,p)$ plane of the flux-like quantity $p
\cdot I(p)$, where $v = [(\lambda/\lambda_0) - 1]c$ is the
classical Doppler velocity, $p$ is the impact parameter in units of
the photosphere radius $R_{\rm phot}$, and $I(p)$ is the emergent specific intensity along $p$.
The sum over $p$ of the quantity $p \cdot I(p)$ at
$v$ corresponds to the total line flux at $v$, shown at the top of
each panel (solid line). Note that for each line, we select a single
H{\sc i} transition to avoid the corrupting effect of line overlap as we near the Balmer jump
(all other bound-bound transitions of H{\sc i} and of other ions are ignored).
A detailed interpretation of these diagrams is given in \citet{DH_05a,DH_05b} and will not be repeated
here. Note the stark contrast between the considerably more extended
regions of emission and absorption (in velocity or equivalently in space dimension) in the
time-dependent models (top row) compared to the very-confined absorption and nearly nonexistant
emission for the corresponding lines in the steady-state counterparts (bottom row).
The same ejecta is covered by both models (same mass shells with same velocities), yet the
Balmer lines show considerably broader profile in the time-dependent case, mimicking
an ejecta with a faster expansion rate {\it at that given time}.

  The time-dependent version of CMFGEN, thus, resolves the discrepancies with
hydrogen Balmer lines obtained with its steady-state version.
While the difficulty with Balmer lines has been documented in the past (see, e.g., UC05),
we find that time-dependence effects conspire to modify the strength of most lines,
not just those of hydrogen. For example, Na{\sc i}\,D undergoes a considerable
broadening and strengthening with the treatment of time dependence, while in the past,
fitting it well required enhancing the sodium abundance by a factor of about four \citep{DH_06a}.
Perhaps the most extreme case in our investigation is He{\sc i}\,10830\AA, which persists
over the entire 6 weeks covered by the baseline model A, and for which, at the last time
in the sequence, we predict a non-trivial absorption strongly blueshifted from line
center (at 1.038$\mu$m, equivalent to -13000\,\kms), and a weak and broad flat-topped
emission\footnote{Chugai et al. (2007) propose that the cool dense shell that forms at the interface
between the SN ejecta and the pre-SN wind is at the origin of a similar, but stronger,
feature in He{\sc i}\,10830\AA, that fits more suitably the observed line
profile \citep{Hamuy_etal_2001,Spyromilio_etal_1991}.
We acknowledge this possibility, and focus here only on the time-dependence effects germane to
the ejecta alone.}.
Such a line profile morphology is synonymous with absorption/emission far above and {\it detached}
from the photosphere, and usually associated with chemical stratification of
the corresponding element or density kinks; here
it stems solely from an {\it ionization stratification}
caused by time-dependence effects. The He{\sc i}\,10830\AA\ line is entirely absent in the steady-state
counterpart model  three weeks after the explosion because, in the
relevant region, helium atoms are once ionized with a probability of only one in 10$^{20-30}$.

\begin{figure}
\epsfig{file=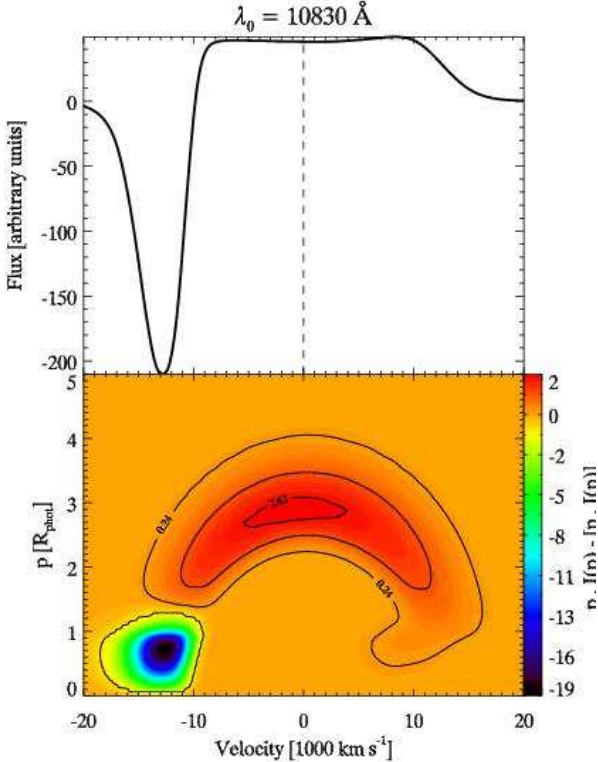,width=8cm}
\caption{
Same as in Fig.~\ref{fig_balmer_pip}, but for the He{\sc i}\,10830\AA\ line
(to enhance the rendering, we subtract $p \cdot I(p)$ from the
continuum-only equivalent, and add a few contours). The line profile is flat-topped, shows a ``detached''
blueshifted absorption, forms further out than H$\alpha$ in the same time-dependent
model, and is completely absent in the equivalent steady-state model.
[See the electronic edition of the Journal for a color version of this figure,
and the text for discussion.]
}
\label{fig_hei_pip}
\end{figure}

\begin{figure*}
\epsfig{file=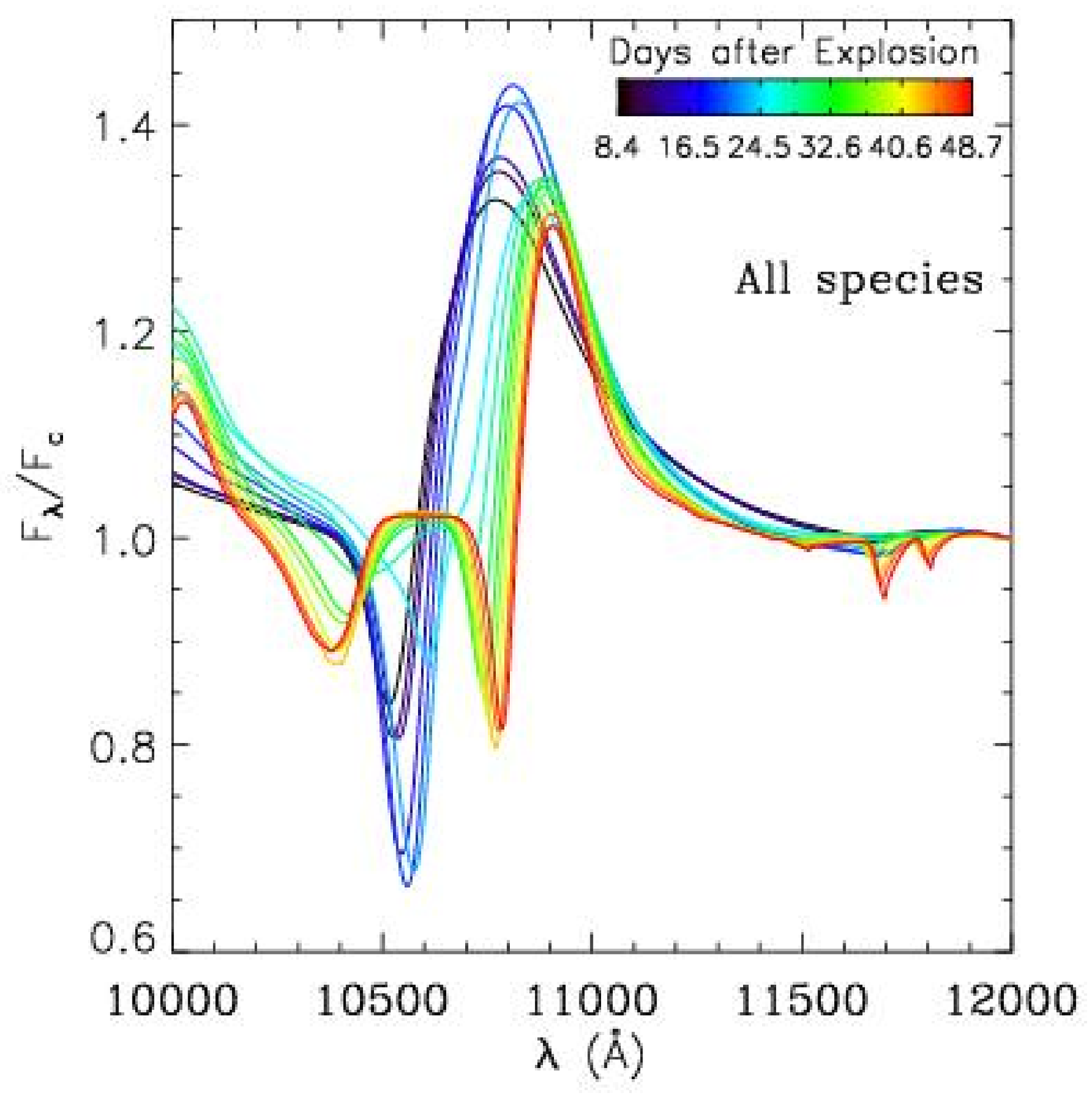,width=8cm}
\epsfig{file=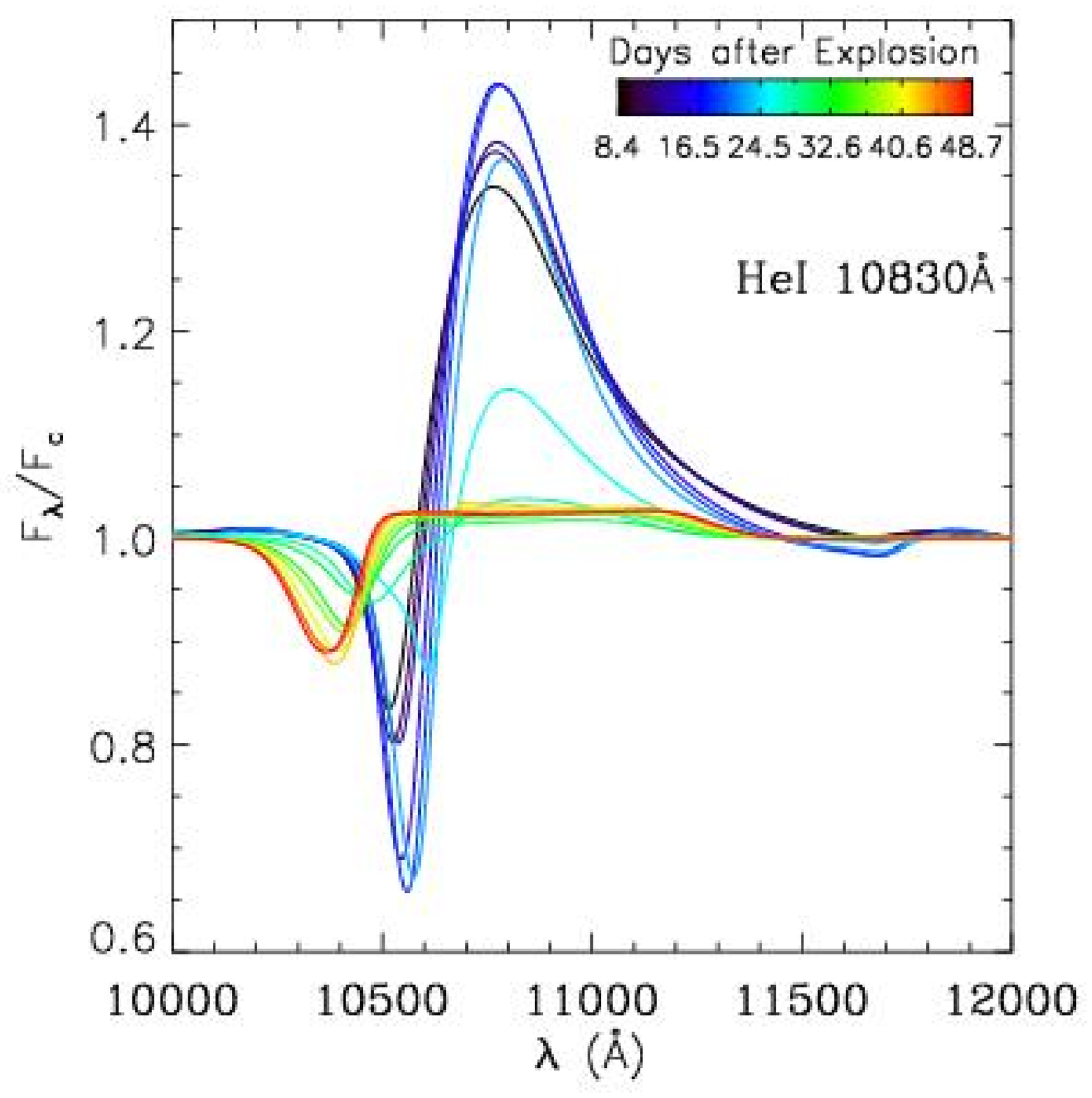,width=8cm}
\caption{
{\it Left}: Time sequence of the rectified synthetic flux over the 10000--12000\AA\ range including
all the bound-bound transitions. Main features are due to P$\gamma$ and He{\sc i}\,10830\AA.
{\it Right}: Same as left, but this time, we only include He{\sc i} bound-bound transitions.
Notice the flat-top and blueshifted absorption of He{\sc i}\,10830\AA\ at late times.
[See the electronic edition of the Journal for a color version of this figure,
and the text for discussion.]
}
\label{fig_montage_hei}
\end{figure*}

In Fig.~\ref{fig_hei_pip}, we reproduce Fig.~\ref{fig_balmer_pip} and
illustrate the absorption/emission sites for He{\sc i}\,10830\AA\
at 48.7 days in the time-dependent version of Model A. Absorption stems from the regions in the
direction of the stellar disk, far from the photosphere which has a velocity
of $\sim$4500\,\kms on that day. For impact rays not intersecting the photodisk, only emission occurs,
again at large distances above the photosphere, and over a small volume. The emission flux is small,
appears as a flat-topped profile, and is thus barely noticeable above the continuum. Of this
He{\sc i}\,10830\AA\ line in the synthetic spectrum, one can only observe the strongly
blueshifted absorption component, the location of the flat-topped emission overlapping
with the much stronger P$\gamma$ line at this late epoch.
This is illustrated differently in Fig.~\ref{fig_montage_hei}, where we show a montage
of synthetic spectra covering the 1--1.2$\mu$m region and the entire evolution computed
for Model A. In the left panel, we show the full synthetic spectrum, i.e., including
all species. In the right panel, we use the converged CMFGEN model at each time in the sequence and
solve the formal solution of the radiative transfer problem by including bound-bound transitions
of helium only. Note how at early times (dark curves) the profiles in the two panels match, indicating that
helium is the main feature contributing in emission and absorption at early times.
After three weeks, the emission is dominated by P$\gamma$ and the He{\sc i} absorption
gives rise to a dip, detached from the P$\gamma$ absorption, and shifting further to the blue (i.e.,
in an opposite sense to the photosphere which recedes to smaller velocities, as well as
to all other line-profile absorptions).

      \subsection{Sensitivity to ejecta properties and epochs}
   \label{comp_with_params}

  Besides the baseline model A discussed above, we have computed
three other time sequences, as described in
detail in \S\ref{sect_models}. Model B and C investigate the effect of the expansion velocity
of the ejecta, model B being also an underluminous event by design. Model D further investigates
the possible (but not expected) time-dependent effects during the first week after explosion.

\begin{figure*}
\epsfig{file=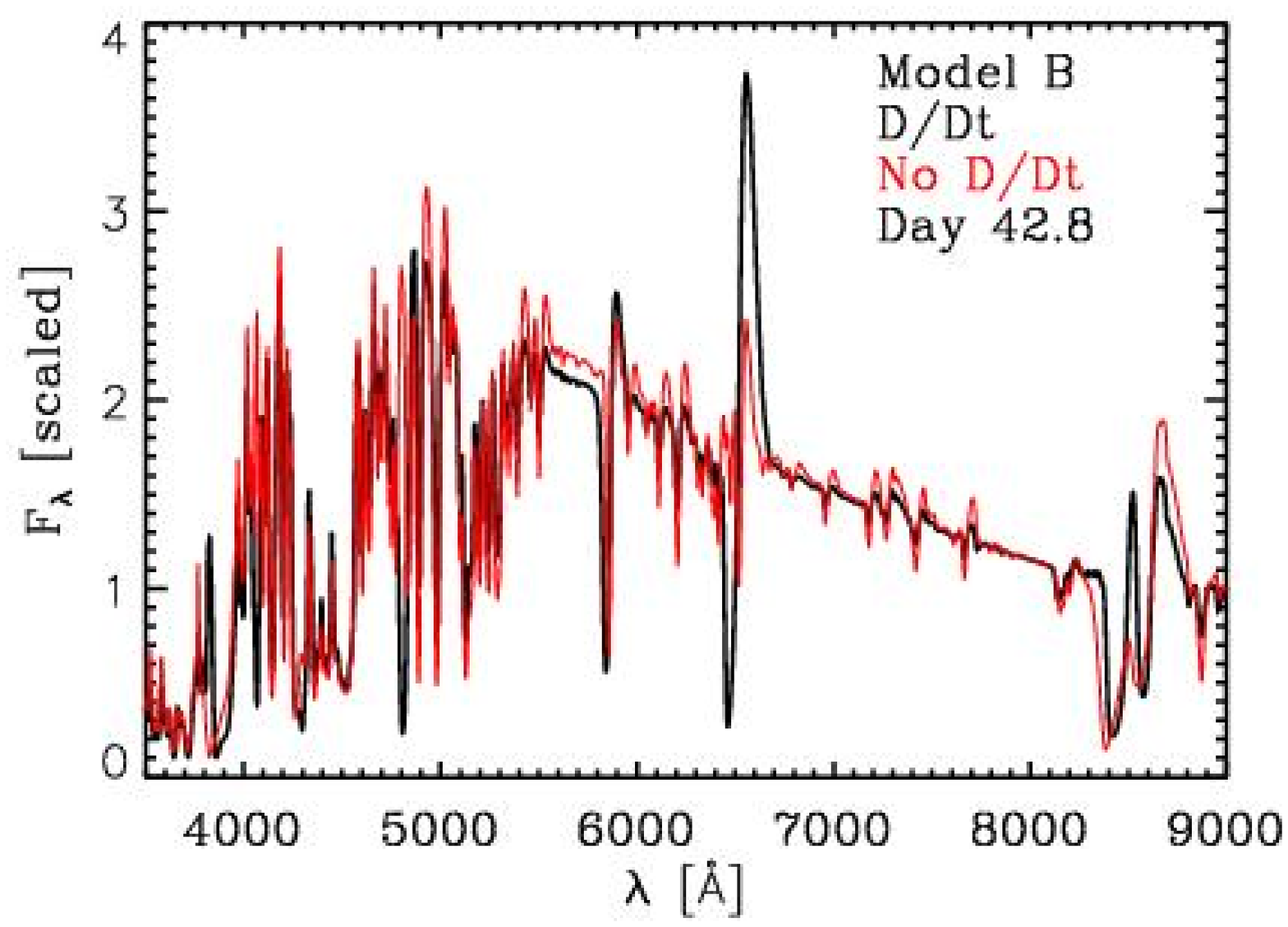,width=8cm}
\epsfig{file=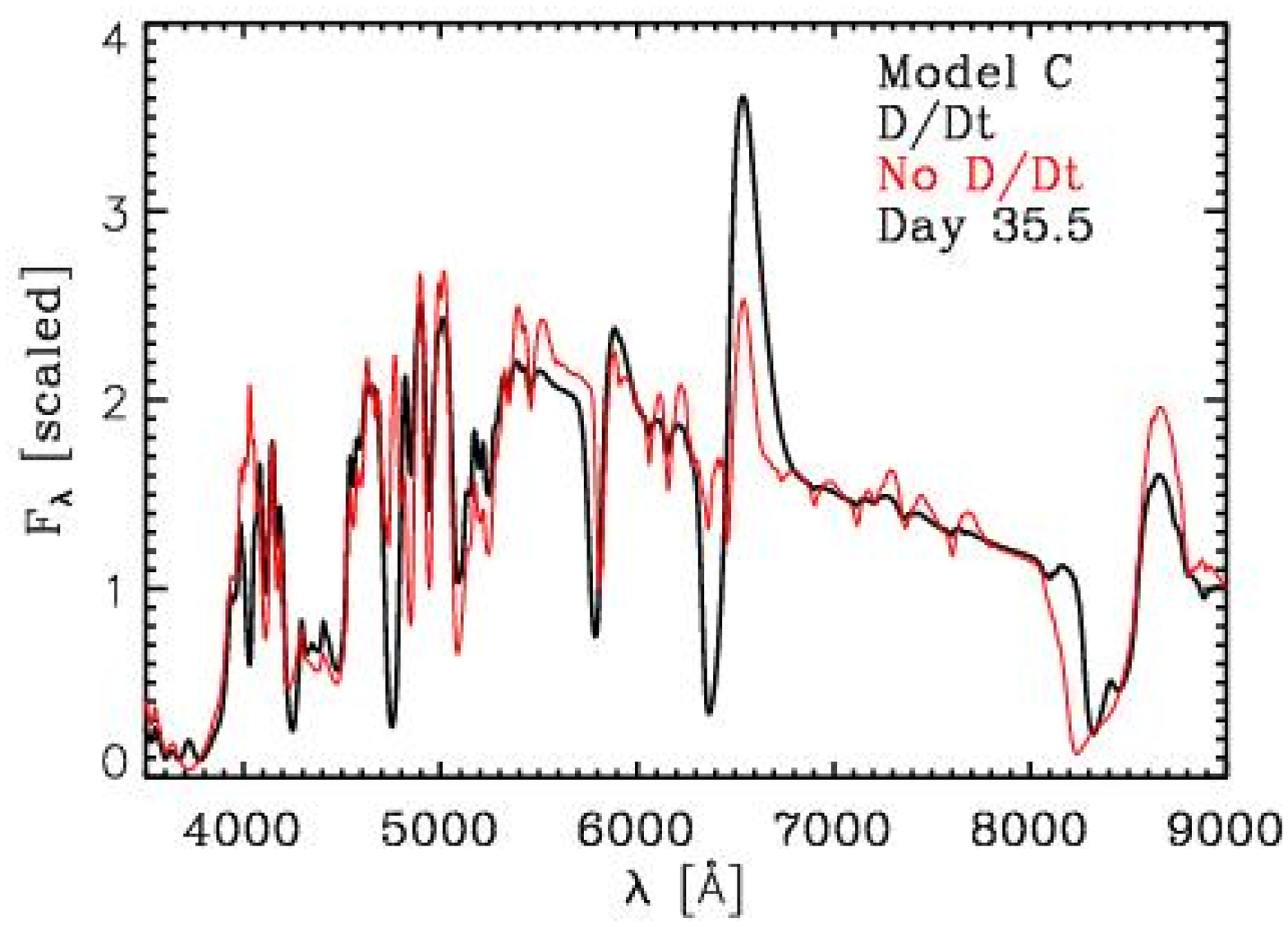,width=8cm}
\caption{
Optical synthetic spectra for the time-dependent (black) and steady-state (red) model B (left;
day 42.8) and model C (right; day 35.5) at the last time in the sequence.
[See the electronic edition of the Journal for a color version of this figure,
and the text for discussion.]
}
\label{fig_spec_modelBC}
\end{figure*}

  We show in Fig.~\ref{fig_spec_modelBC} a comparison between optical synthetic spectra computed for
the time-dependent (black) and the steady-state (red) Model B (left) and Model C (right), both
at the last time in the sequence.
  Despite their different expansion velocities, models B and C show very similar time-dependence
effects, with the most noticeable features being the same as for model A over the
same time range: Hydrogen Balmer lines are strengthened, and so are those
of the Paschen and the Brackett series (not shown here but see Fig.~\ref{fig_spec_modelA} for a guide),
as well as Na{\sc i}\,D. Here again, and by contrast, the Ca{\sc ii} multiplet at $\sim$8500\,\AA\ is weakened.
Importantly, given that these models correspond to the conditions from underluminous to standard,
and from slow to fast ejecta, germane to SNe 1999br/2005cs and 1999em/2006bp, time-dependent CMFGEN models
systematically predict strong spectroscopic effects.
In an absolute sense, these modulations still correspond to ejecta
that are fast and tenuous, where the expansion time scale is comparable to
the magnitude of important rates (like that of recombination), affecting strongly the ionization
equilibrium at and above the photosphere.

\begin{figure}
\epsfig{file=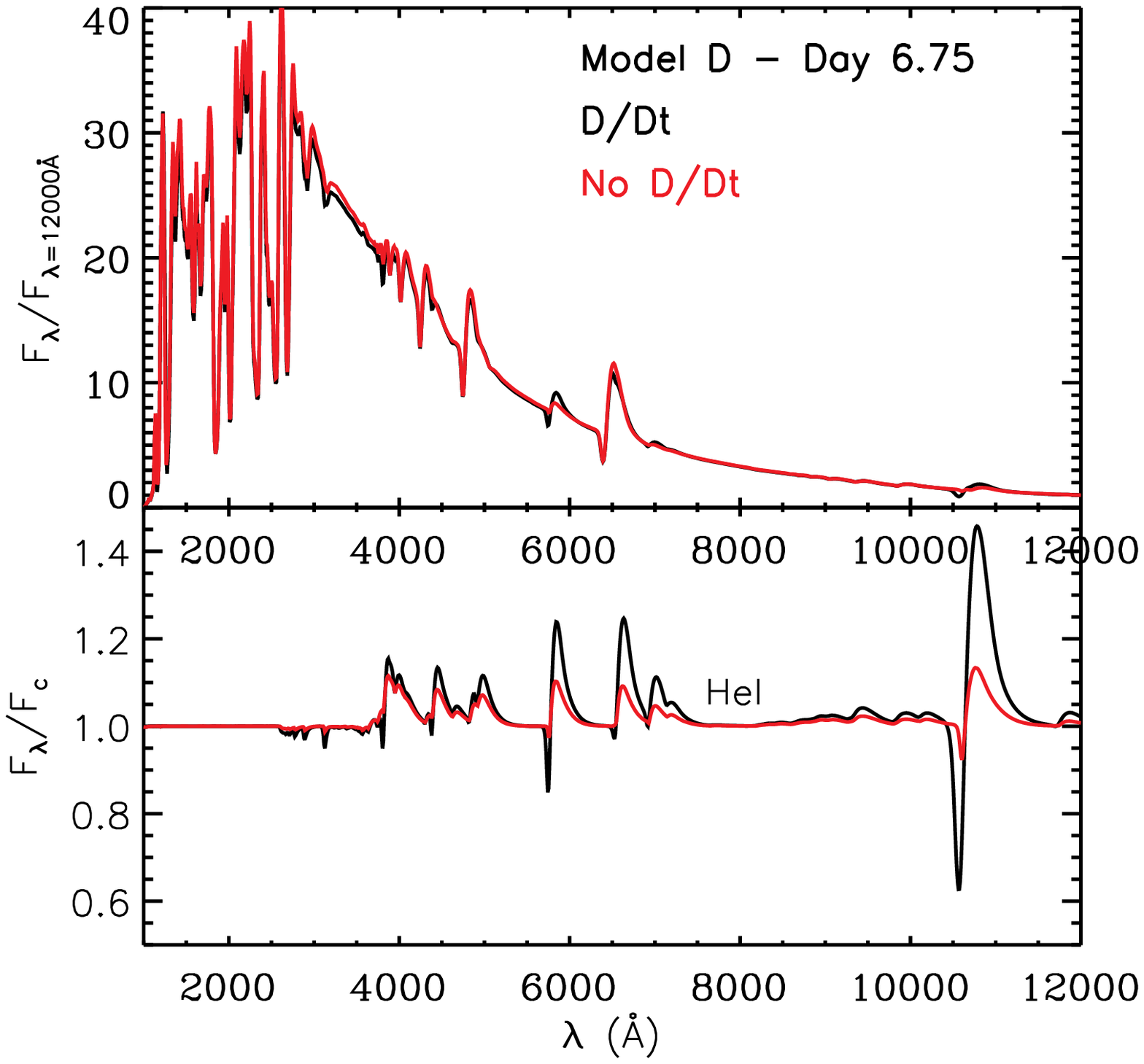,width=8cm}
\caption{{\it Top:} Comparison of synthetic spectra obtained for model D at 6.8 days
for the time-dependent (black) and steady-state (red) approaches.
{\it Bottom:} Same as top, but for the synthetic rectified spectra accounting
only for He{\sc i} bound-bound transitions. No time-dependent effect in the emergent
light is seen for the other species treated in the calculation.
(See text for discussion.)
}
\label{fig_comp_DDT_noDDT_early}
\end{figure}

 Turning to Model D in Fig.~\ref{fig_comp_DDT_noDDT_early} we see that the differences,
at 6.7 days, between the time-dependent and steady-state
model SEDs are subtle (bottom panel). The continuum SED is essentially identical and
noticeable differences are only visible in the He{\sc i} lines, particularly at 5875\AA\
and at 10830\AA.
Both of these are a factor of about 2--3 stronger in the time-dependent model.
In the past, reproducing the strength of optical He{\sc i} lines with a typical BRG/RSG
surface helium abundance has been a challenge (Eastman \& Kirshner 1989),
although this difficulty disappears with a full non-LTE treatment \citep{DH_05a}. We now find that
time dependence introduces a further tuning, making optical He{\sc i} lines persist over a
longer time span than is predicted by (non-LTE) steady-state models.

 \section{Comparison with a sample of observations}
 \label{comp_with_obs}

  The motivation for the present work was to find an explanation, and a cure,
for our inability with steady-state CMFGEN models to reproduce the
strong H$\alpha$ line at the recombination epoch.
Despite all attempts, we were unable to reproduce the H$\alpha$ line strength as soon
as hydrogen recombination in the Type II SN ejecta takes place \citep{DH_06a}.
We reported this problem for SN 1999em, but it is in fact a generic feature of {\it all}
steady-state CMFGEN models at a few weeks after explosion. Interestingly, observations
at the corresponding epoch reveal that H$\alpha$ is instead the {\it strongest} line
in the optical spectrum of Type II SNe, across the whole range of objects, from
the peculiar SN 1987A, to the underluminous slowly-expanding SN 1999br,
to the more standard SN 1999em (Type II-P).
These SNe are understood to stem from core-collapse explosions
of supergiant stars, and are characterized by a wide range of explosion energies (velocity
distribution versus Lagrangean mass), nickel yields, and bolometric luminosity. The mechanism
that can solve the present H$\alpha$ discrepancy must operate at the
hydrogen recombination epoch and be effective in all Type II SN ejecta.

Specifically, it is unlikely that, in general, non-thermal ionization/excitation by $^{56}$Ni is the
missing mechanism in our former approach \citep{DH_06a} since SNe events like 1999br or 2005cs,
which both boast a strong H$\alpha$ line at the hydrogen recombination epoch, are also
characterized by vanishingly small $^{56}$Ni yields \citep{Pastorello_etal_2004,T06,TV06,Pastorello_etal_2006}.
Furthermore, the H$\alpha$ discrepancy appears at times when the photosphere still moves at
4000-5000\,\kms, velocities out to which $^{56}$Ni is not expected to
be mixed routinely \citep{Fryxell_etal_1991,Kifonidis_etal_2000,Kifonidis_etal_2003}.
The H$\alpha$ problem
in SN 1987A, for example, is already visible at just a few days after explosion, while Cobalt lines,
testifying for the presence of radioactive material in the photosphere layers,
were first seen at about 45 days after explosion (see, e.g., Pinto \& Woosley 1988),
significantly later (but much earlier than anticipated by the theory at the time SN 1987A went off).

\begin{figure*}
\epsfig{file=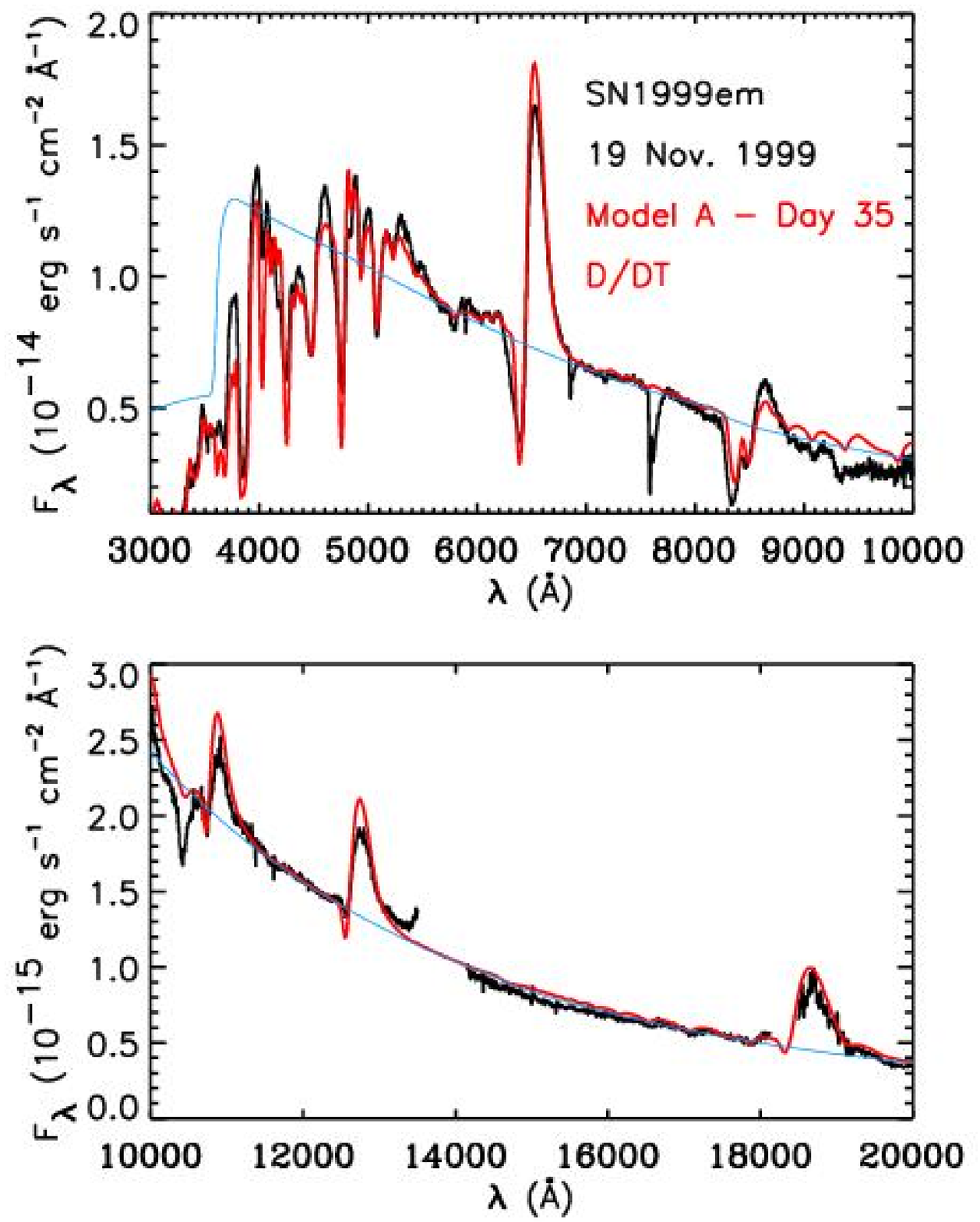,width=8cm}
\epsfig{file=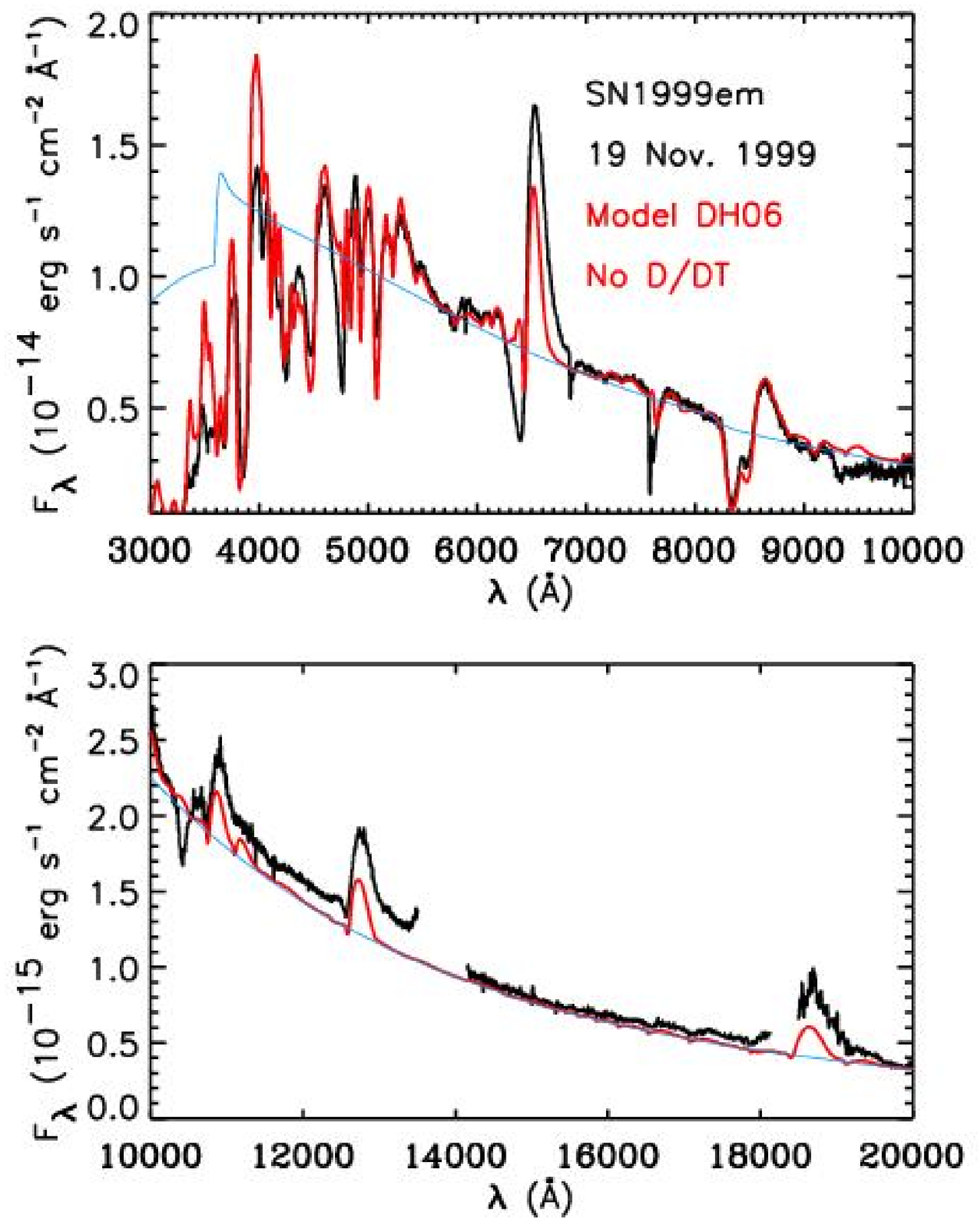,width=8cm}
\caption{
{\it Left:} Illustration of the match between observations (black) of SN 1999em on the
19th of November 1999 in the optical range (top; Leonard et al. 2002) and
on the 18th of November 1999 in the near-IR range (bottom; Hamuy et al. 2001),
with the reddened synthetic spectrum (red; blue: continuum-only)
for the time-dependent model A at 35 days after explosion.
We use the extinction law of Cardelli et al. (1988), with $E(B-V)=0.1$,
and normalize the synthetic flux to the observed flux at 6100\AA,
with a further 15\% flux reduction in the near-IR (note the one day time
difference between the two spectra).
{\it Right:} Same as left, but this time using the equivalent steady-state model.
[See the electronic edition of the Journal for a color version of this figure,
and the text for discussion; see also \S3.2.7 and Fig.~7 in Dessart \& Hillier (2006a).]
}
\label{fig_99em_1911}
\end{figure*}

We start the comparison with optical and near-IR observations of SN 1999em on the
18th-19th of November (Leonard et al. 2002, Hamuy et al. 2001). These observations were
modeled by \cite{DH_06a} and we reproduce their Fig.~7 in the right panel of
Fig.~\ref{fig_99em_1911}. In the left panel, we fit the observations with the synthetic spectrum
computed for model A on day 35, and include the near-IR range to emphasize the
concomitant strengthening of the Paschen lines in the time-dependent model.
Time-dependence effects are thus not limited to H$\alpha$, nor to hydrogen Balmer lines exclusively,
but operate as well on the Paschen series, and, even more generally, affect lines of
all species.
Note also that the predicted He{\sc i} line at 10830\AA\ is too weak to fit the absorption
feature at 1.03$\mu$m, which, despite our neglect of C{\sc i} contributions \citep{DH_06a},
may require the additional contribution from the Cold Dense Shell at the ejecta/pre-SN-wind
interface \citep{Chugai_etal_2007}.
Note that the time-dependent model, unlike the steady-state model shown on the right, was not
tuned to fit the observations. Hence, the adopted luminosity evolution may not be fully adequate
for this event.
An additional issue is that model A started at one week after explosion, while we demonstrated
clearly in \S\ref{effects_on_light} the strong time-dependence effects occuring for He{\sc i}
even at such early times, effects that will carry over to subsequent times and that may therefore
be underestimated in these model sequences.

\begin{figure*}
\epsfig{file=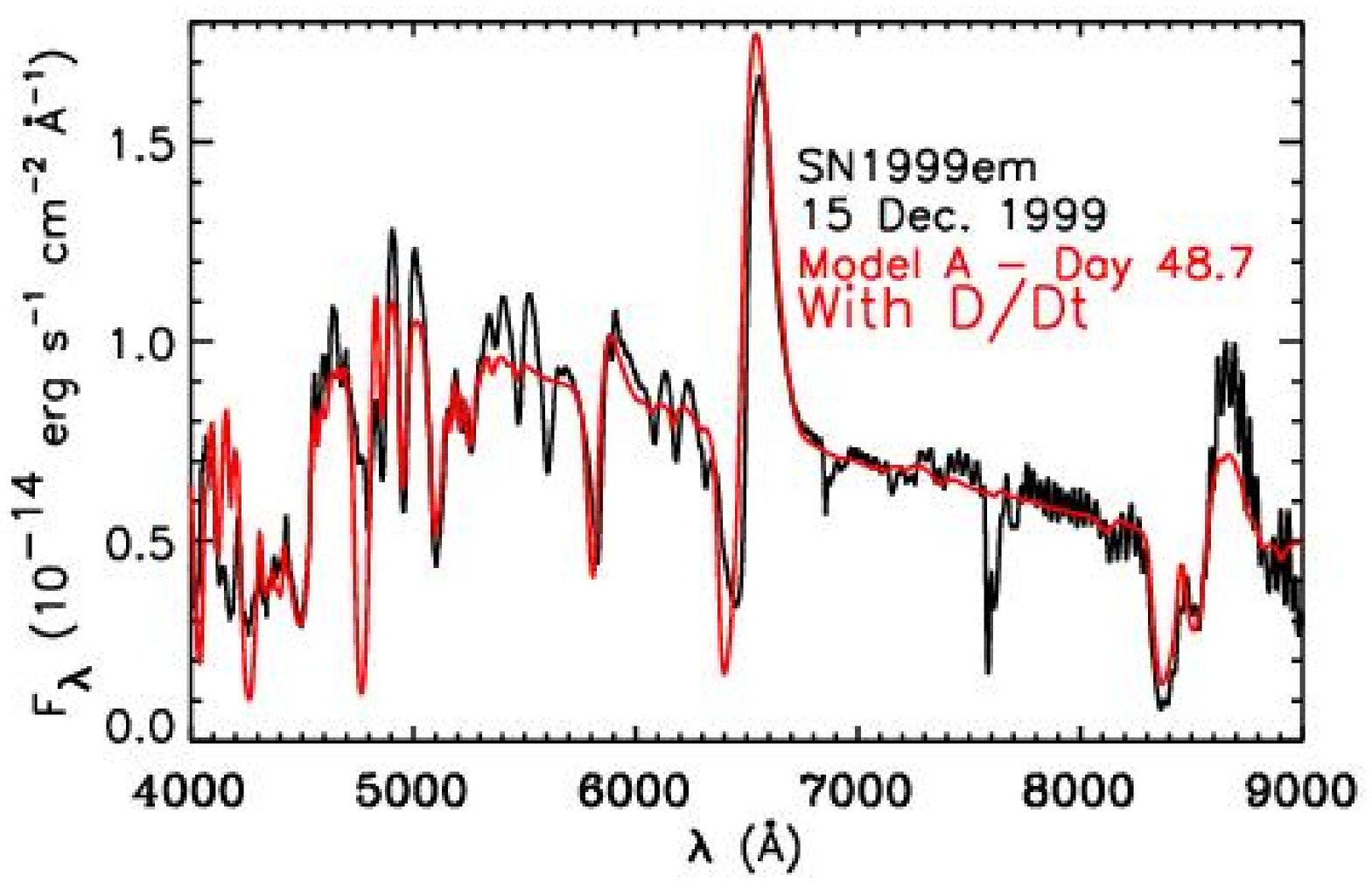,width=8cm}
\epsfig{file=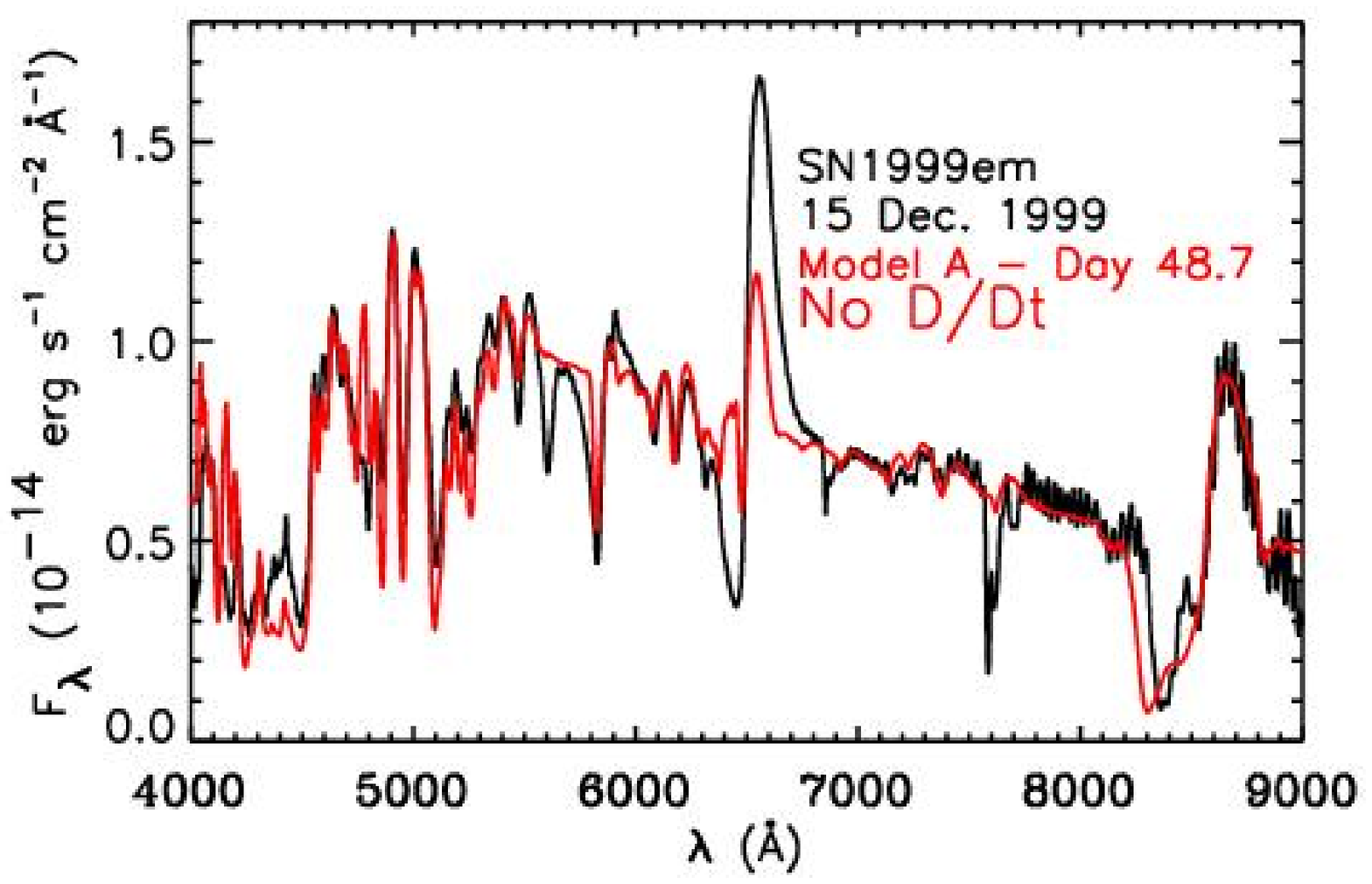,width=8cm}
\caption{
Same as Fig.~\ref{fig_99em_1911}, but for the observations of SN 1999em on the
15th of December 1999 (black; Leonard et al. 2002) and using the time-dependent model A
at 48.7 days after explosion.
[See the electronic edition of the Journal for a color version of this figure,
and the text for discussion.]
}
\label{fig_99em_1512}
\end{figure*}

In Fig.~\ref{fig_99em_1512} we compare synthetic spectra for Model A at 48.7 days
after explosion, with allowance for time dependence (left panel) or assuming steady-state (right panel),
with the observations of SN 1999em taken on the 15th of December 1999 (Leonard et al. 2002).
The SED has changed significantly from that seen on the 19th of November 1999.
There is more flux depletion in the
UV, more metal line-blanketing in the optical, and H$\alpha$, Na{\sc i}\,D, and
Ca{\sc ii}\,8500\AA\ have strengthened. The steady-state model (right panel) reproduces well the overall
shape of the SED, but what was a slight discrepancy in fitting the H$\alpha$ line at earlier epochs
now becomes a major discrepancy. By contrast, the time-dependent model reveals a strong and broad
H$\alpha$ line that closely matches the observed profile. The same agreement is reached for Na{\sc i}\,D.
A number of metal species are missing (scandium, barium) and the CNO abundances would also need
adjustment (enhancement of carbon and oxygen abundances, and depletion of nitrogen; Dessart et al. 2007)
but overall, the time-dependent model allows a much more satisfactory fit to observations
and in particular H$\alpha$, which is the defining feature of Type II SN spectra.

Note that the alternatives, an enhancement in turbulent velocity  or flattening of the density
distribution, do not improve the agreement with observations. The former has a weak or negligible
effect at late times, while the latter leads to a large enhancement of the Ca{\sc ii}\,8500\AA\
line strength, in stark contradiction with observations. The key issue here is that one should
not view the recombination rate as scaling with the density squared, but rather as
the ion density times the free-electron density. Once the material has fully recombined,
the absence of free-electrons and ions completely shrinks the recombination rate,
and H$\alpha$\ becomes only weakly sensitive to modulations in the outer mass density distribution.

\begin{figure*}
\epsfig{file=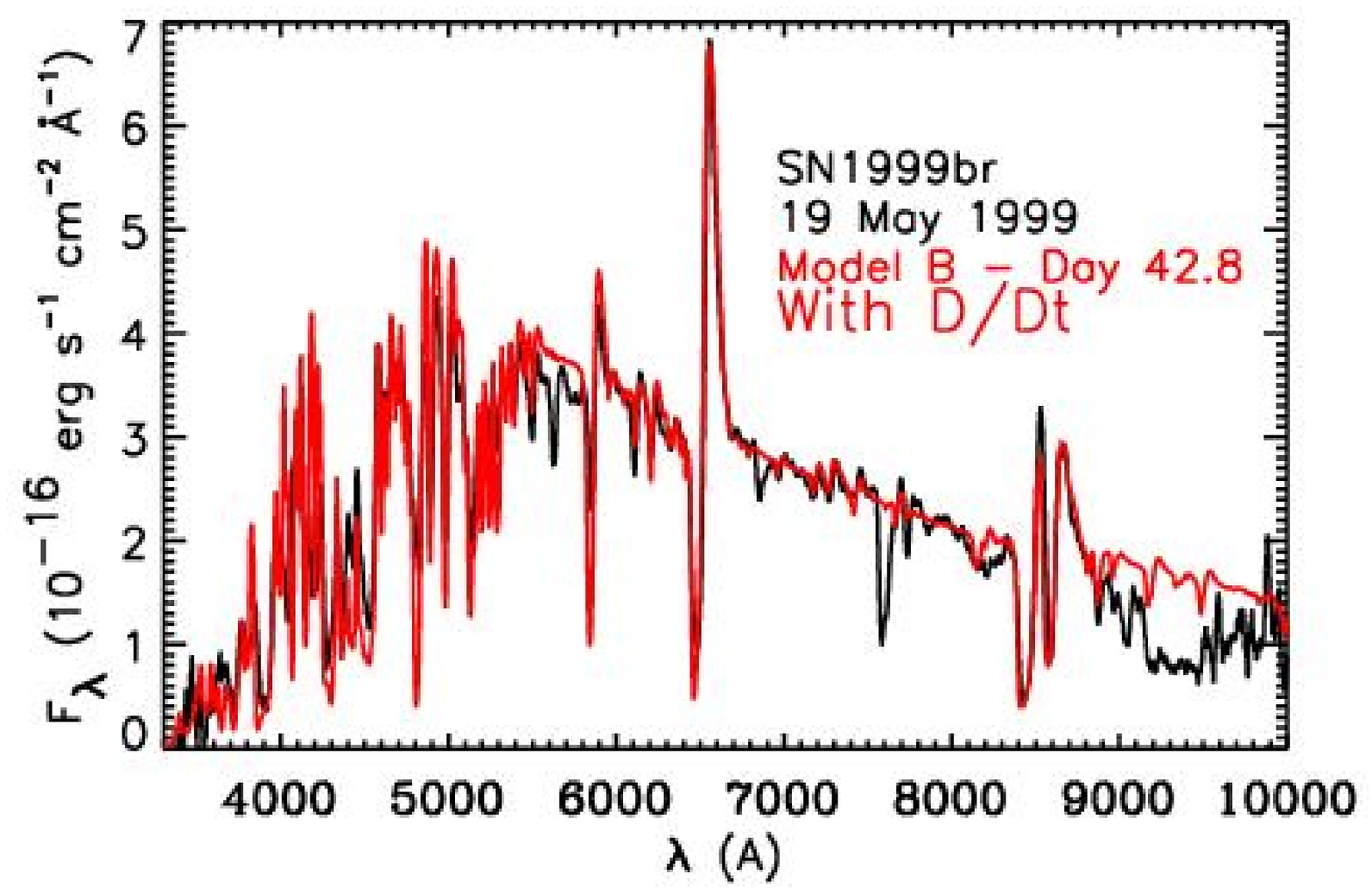,width=8cm}
\epsfig{file=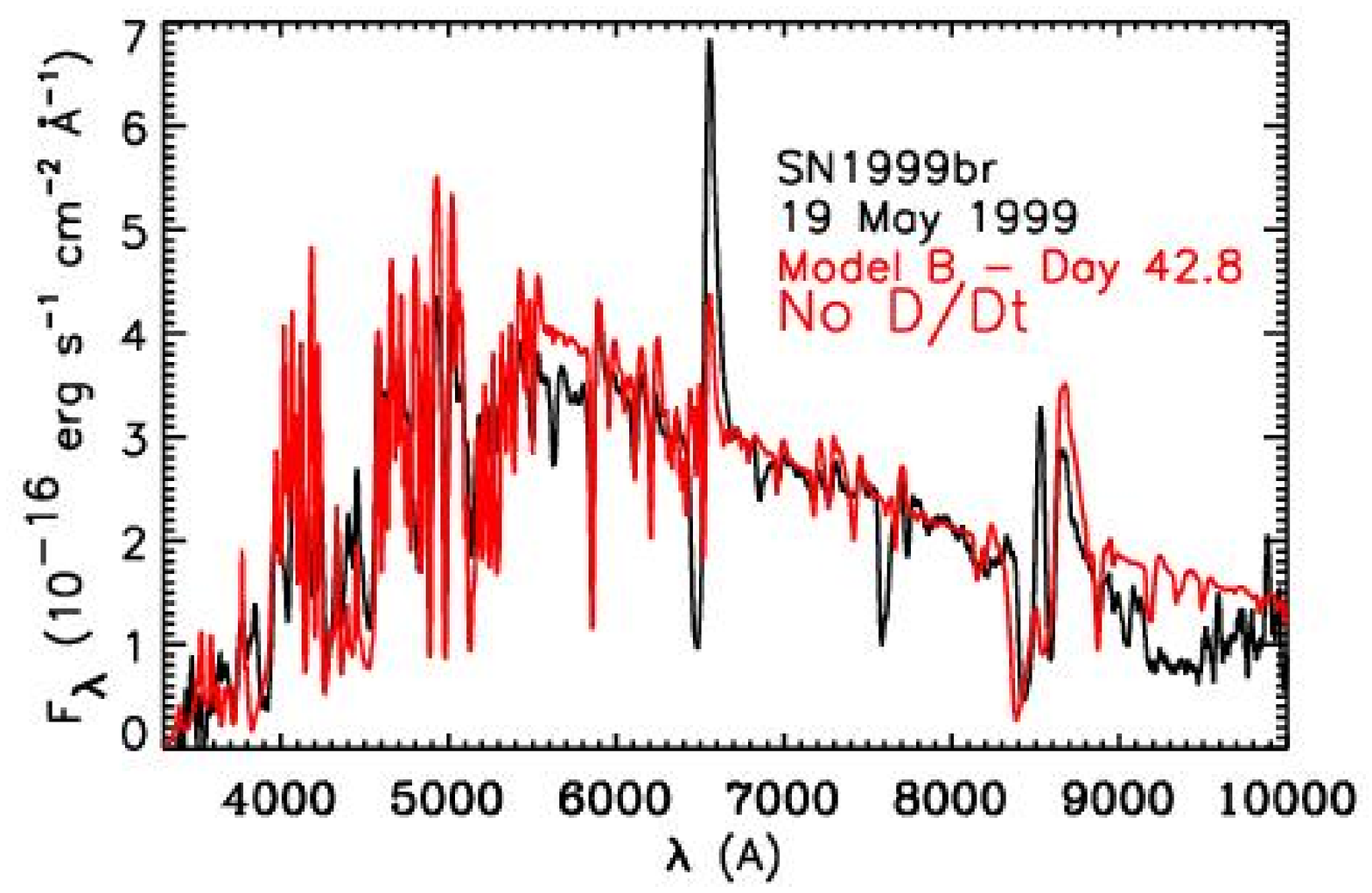,width=8cm}
\caption{Same as Fig.~\ref{fig_99em_1911}, but for the observations of the underluminous
event SN 1999br (black line; Pastorello et al. 2004) and using the time-dependent model
B at 42.8 days after explosion.
Synthetic spectra are reddened using the extinction law of Cardelli et al. (1988)
with $E(B-V)=0.03$.
[See the electronic edition of the Journal for a color version of this figure,
and the text for discussion.]
}
\label{fig_99br_19may}
\end{figure*}

We finally show in Fig.~\ref{fig_99br_19may} a comparison of Model B at 42.8 days after explosion,
including time dependence (left panel) or assuming steady state (right panel),
with the observations of SN 1999br on the 19th of May 1999 (Pastorello et al. 2004).
The time-dependence effects are similar to those described for SN 1999em
on the 15th of December 1999, but they bear more strongly on the role of time dependence since,
for the underluminous
low-nickel-yield SN 1999br, non-thermal ionization/excitation is very unlikely
to affect in any way H$\alpha$ emission. Here again, time dependence effects
lead to strong Balmer (and Paschen) lines, as observed, without recourse to extra
ionizing/exciting sources or changes in the density profile.
Furthermore, this shows that time-dependent effects are present in a wide variety of SN ejecta,
even those characterized by a slow expansion rate.

\section{Implications}

   \subsection{Abundance determinations}
   \label{implications_on_chemistry}

The persistence of He{\sc i}\,10830\AA\ at 6 weeks after explosion in Model A
suggests that the so-called detached absorption/emission lines are not exclusively associated
with a stratification of abundances, with specific ions present at certain heights or
mass shells in the ejecta.
Here, the detached absorption/emission from He{\sc i}\,10830\AA\ stems purely from
a time-dependence effect, and thus does not {\it in principle} require chemical
stratification.
Time-dependence effects can alter our interpretation of
chemical stratification based on line profile measurements, e.g., velocity at maximum absorption.
This is supported further by the time-dependence effects we see on the profile shape
of strong lines, such as, for example, the hydrogen Balmer lines or Na{\sc i}\,D. In the time-dependent
model A at 48.7 days, we measure from the synthetic spectrum that included only the bound-bound transition
of the relevant species a velocity at maximum absorption of -7500\,\kms\ for H$\alpha$, -5800\,\kms\
for H$\beta$, -4900\,\kms\ for H$\gamma$, -4000\,\kms\ for Na{\sc i}\,D (at 5890\AA), while in the same order,
measurements performed on the synthetic spectrum for the steady-state model give
-3900\,\kms, -3500\,\kms, -3400\,\kms, and -3200\,\kms. Differences between these values
are greater than those in photospheric velocity between the two models, which, respectively, are
4130\,\kms\ and 3510\,\kms. Time-dependence effects operate in a non-linear fashion, affecting
lines and the continuum with a different magnitude, something to be expected since the
continuum is mostly sensitive to the ionization state (a global quantity) while line profiles
are affected by many parameters, such as the relevant ion level populations,
the electron density, the radiation field etc. over a broad region in the SN ejecta.

Tomographic techniques, which aim at constructing the distribution of elements with mass
in SN ejecta through the assumption of homologous expansion and kinematic measurements on
(or fits to) line profiles \citep{Stehle_etal_2005}, may be compromised through the
neglect, in the solution to the radiative transfer problem, of time-dependent terms in
the statistical and radiative equilibrium equations.
In the future, we will investigate such effects on inferences of chemical stratification
in SN ejecta.

   \subsection{Correction factors}
   \label{implications_on_xi}

  Type II SNe offer an attractive means to determine distances to galaxy hosts,
and ultimately, constrain distances in the Universe. The so-called Expanding
Photosphere Method (EPM; Kirshner \& Kwan 1974) has been used and improved
upon to yield distances to a few nearby Type II SNe (Eastman \& Kirshner 1989;
Schmidt et al. 1994; Eastman et al. 1996), although with a moderate level of success
for SN 1999em for which the EPM distance (Hamuy et al. 2001; Leonard et al. 2002a; Elmhamdi et al. 2003)
underestimates by 50\% the Cepheid distance (Leonard et al. 2003) to the galaxy host.
Using the updated correction factors of  Dessart \& Hillier (2005b) partially cures this
discrepancy, while even more consistent results require the use of tailored models, with their
combined correction factors and color temperatures (Dessart \& Hillier 2006;
Dessart et al. 2007).

Originally, the correction factors were introduced to account for the flux dilution that
stems from the large electron-scattering optical depth contribution of the photosphere.
However, as noted by Dessart \& Hillier (2005b), they also correct for the
(strong) deviations from a blackbody distribution arising from the
UV and optical line blanketing, particularly at later times.

\begin{figure*}
\epsfig{file=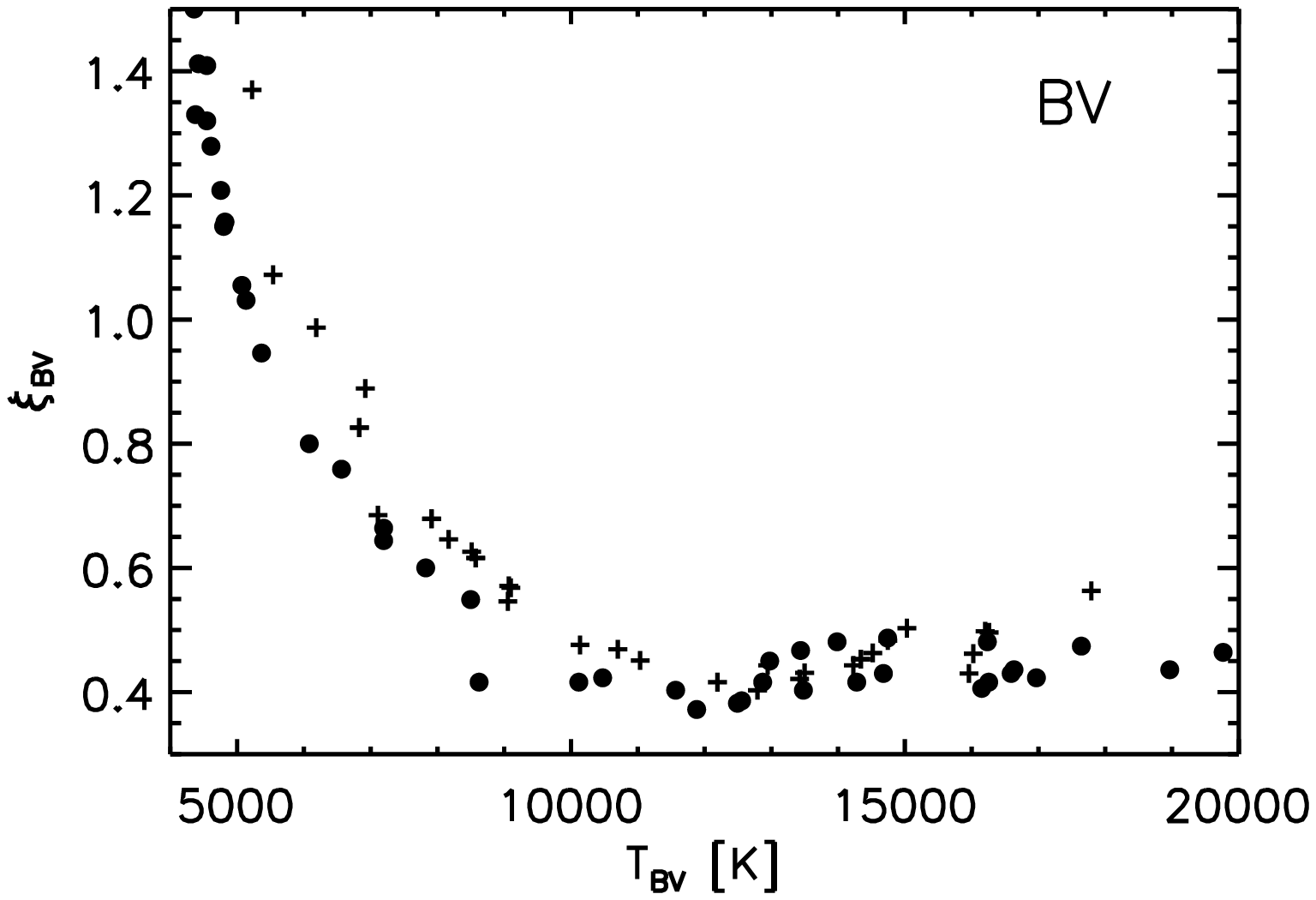,width=5.5cm}
\epsfig{file=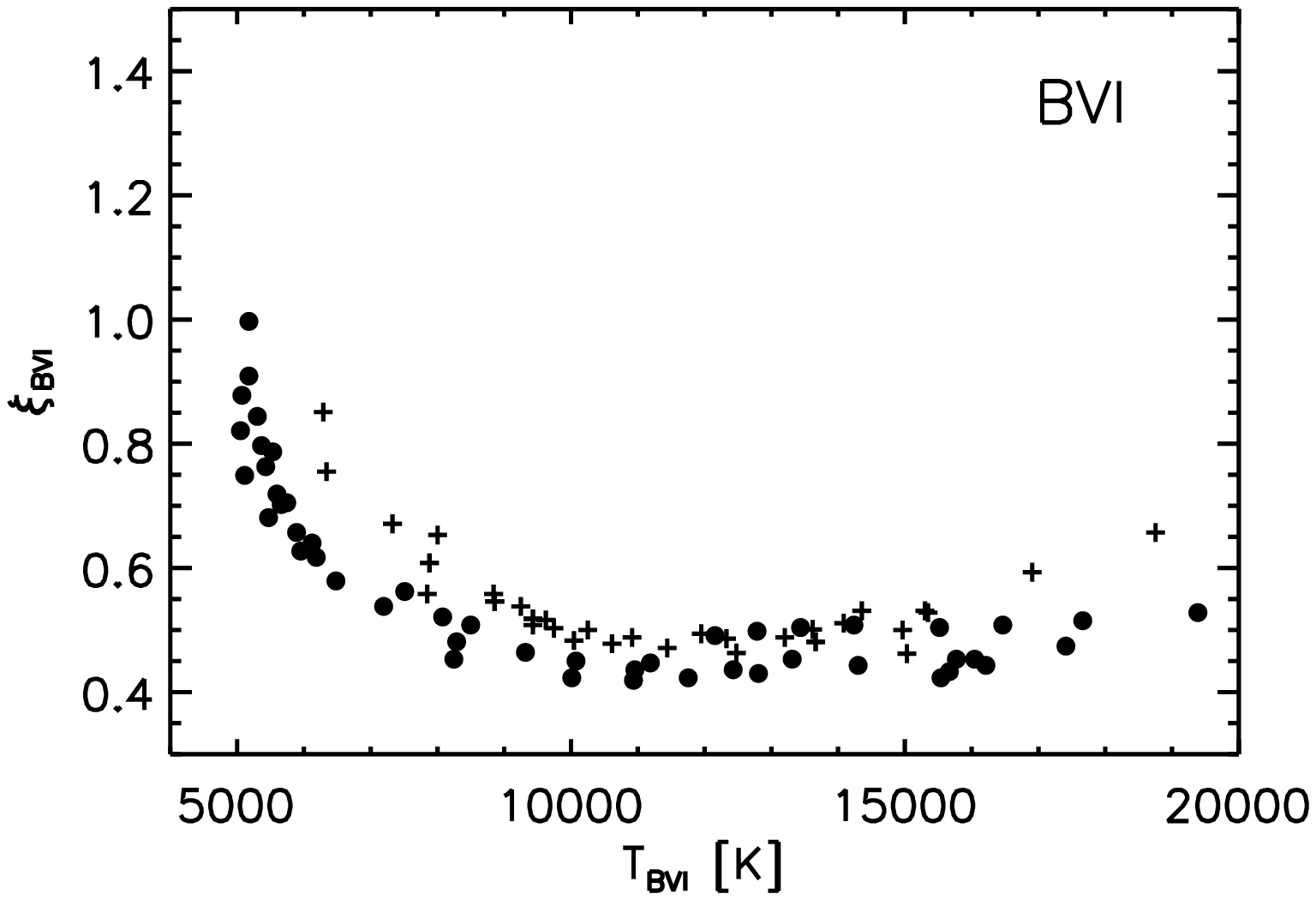,width=5.5cm}
\epsfig{file=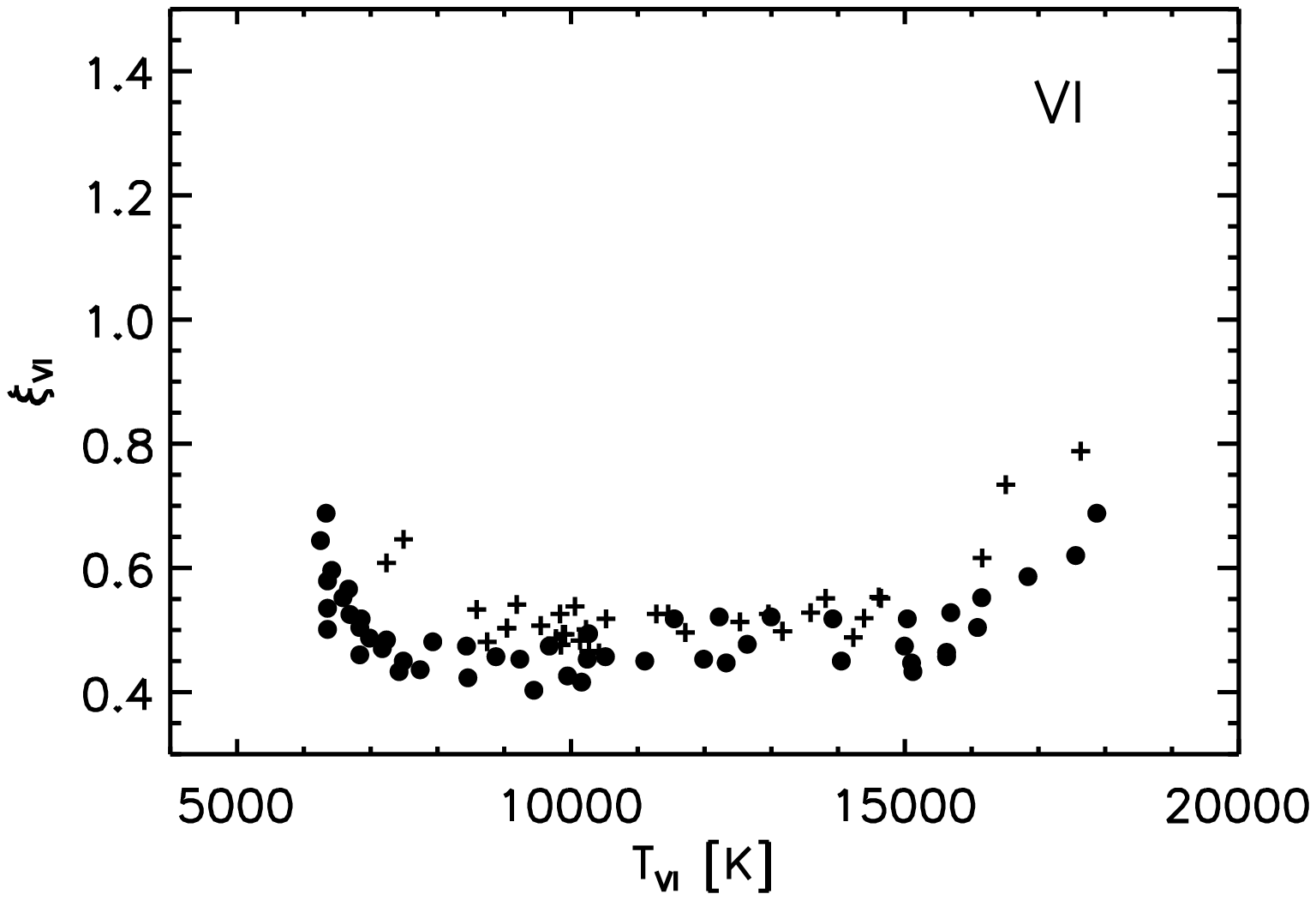,width=5.5cm}
\caption{
Correction factors $\xi_S$ versus corresponding color temperature $T_S$ for the
51 time-dependent models computed in the present study (dots), and shown for
bandpass sets $BV$ (left), $BVI$ (middle), and $VI$ (right).
For comparison, we also overplot crosses for the $\{\xi_S$,$T_S\}$ couples obtained with
steady-state CMFGEN models for SN 1999em \citep{DH_06a}, SNe 2005cs and
2006bp \citep{Dessart_etal_2007}.
There is broad agreement between the time-dependent and steady-state predictions,
although at low temperatures there is a tendency for the time-dependent correction
factors to lie at the lower edge of the distribution for the steady-state models.}
\label{fig_epm}
\end{figure*}

  Here, we investigate whether time-dependent models yield correction factors and associated
color temperatures that depart systematically from steady-state models. Time dependence affects flux dilution
by enhancing the electron density (and thus the magnitude of flux dilution) at all times,
relative to a steady-state
approach. However, it  also affects the SED directly, both for the continuum (weakly) and for lines
(strongly). We show in Fig.~\ref{fig_epm} the collection of correction factors versus
color temperature for each set $BV$ (left), $BVI$ (middle), and $VI$ (right),
for all models computed in the present study (51 models, shown as dots).
For comparison, we overplot correction factors and associated color temperatures
computed with steady-state CMFGEN models (37 in total, shown as crosses)
for the study of SN 1999em \citep{DH_06a},
as well as that for SNe 2005cs and 2006bp \citep{Dessart_etal_2007}.


At temperatures above 10,000K  we find no sizable systematic difference in correction
factors computed with the time-dependent and steady-state approaches, which supports
the use of the {\it steady-state} models of Type II SNe for distance determinations. At lower
temperatures, particularly at temperatures less than 7000K, the new correction factors
are somewhat smaller, and tend to lie at the lower edge of the distribution of correction factors
calculated from steady-state models. They are also closer to those of Hamuy et al. (2001).
The large scatter from model to model suggests that a tailored analysis is nonetheless
preferred over the use of analytical correction factors.
The departures we observe at low temperatures are not surprising ---
this is a regime where the time-dependent terms have a very large effect on spectral features,
and a regime which we have tried to avoid in our EPM distance determinations.
The inclusion of the time-dependent terms
will allow us to better model the long-term evolution of SN spectra, and hence to extend the
time baseline over which the EPM can be used. Further, the inclusion of these time-dependent
terms is also important if we are to reduce systematic errors, and realize the fullest potential of
the EPM technique.

%

   \subsection{Polarization}
   \label{implications_on_pol}

  Through its effect on the ionization equilibrium of Type II SN ejecta, time dependence
leads to an enhancement in the ejecta electron density.
The ejecta optical depth falls off more rapidly than
$1/t^2$, but not as fast as it would for a steady-state configuration (Fig.~\ref{fig_tau_es}).
Deeper layers should thus be revealed with a delay and the nebular phase be entered later
in nature compared to steady-state predictions.

Moreover, the approximate linear scaling of
the degree of polarization with the electron-scattering optical depth \citep{Brown_McLean_1977}
suggests that, for a given degree of asymmetry, higher polarization measures will result from
theoretical models generated assuming time-dependence, rather than stationarity.
In Type II-P SNe, polarization is usually inferred to be at the 0.1\% level during the
plateau phase, and, although late time polarization measures are rare, SN 2004dj indicated
that polarization can increase dramatically (here to $\sim$1\%) when the core is revealed
\citep{Leonard_etal_2006}. Understanding what controls the ionization at such late times,
in particular the respective roles of time dependence and radioactive decay from $^{56}$Co,
is therefore critical for a proper interpretation of the degree of polarization from SN
observations and the link to the asymmetry of the explosion.

\section{Conclusion}

   We have presented an extension to the non-LTE model atmosphere code CMFGEN
to treat time-dependent terms in the statistical and radiative equilibrium equations.
We use implicit first order differencing, and the resulting equations are solved by a
partial linearization technique in a manner virtually identical to that used to solve the steady state
equations \citep{HM_98}. We confirm the findings of Utrobin \& Chugai (2005), which were
related to H$\alpha$ and Ba{\sc ii}\,6142\AA\  at a few days after explosion in the
spectrum of SN 1987A, that the time dependent terms are important for the analysis
of Type II SN spectra.


We find that the inclusion of time-dependence terms in the statistical equilibrium equations
produces ejecta, in Type II SN, that are systematically over-ionized relative to steady-state models.
Spectroscopically, the associated changes in level populations and optical depths
alter the line profiles of all species throughout the spectral range.
Qualitatively, lines appear in general stronger and broader, as is seen for
the Balmer, the Paschen, and the Brackett series of hydrogen or for Na{\sc i}\,D. Exceptions
(e.g. Ca{\sc ii}\,8500\AA) arise when the altered ionization inhibits the necessary recombination
of a given ion (e.g., Ca$^{++}$). Because of optical depth effects, the inclusion of
time-dependent terms in the statistical equilibrium equations can even influence the ionization state of
the gas in regions where the classic recombination times-scale is much shorter than the
flow time-scale. Surprisingly, we predict the presence of He{\sc i}\,10830\AA\
6 weeks after the explosion, naturally caused by the frozen, and full, ionization of the outer
ejecta. The resulting detached blueshifted absorption and flat-topped emission profile
is supported by observations, although the magnitude of the absorption component
may require the additional contribution from the cold dense shell that can form at the
interface between the SN ejecta and the pre-SN wind \citep{Chugai_etal_2007}.

The effects of time dependence observed stem from the energy gain that follows
changes in ionization and excitation, and from the similarity in recombination and expansion
time-scales (crucial at large distances above the photosphere). The recombination
time-scale can be lengthened by optical-depth effects, and in some cases this may
allow time-dependence effects to be important even at the photosphere.
The present study indicates that neglecting time dependence may compromise the
results of quantitative analyses of Type II SN spectra, affecting the
inferred ejecta ionization (and its origin) or chemical abundances.
In particular, time dependence offers a natural means to sustain strong and broad
lines at late times, as observed in {\it all} Type II SN spectra, without invoking
radioactive contributions. Such non-thermal excitation/ionization is expected to be
relevant at the photosphere after a few months, but not as soon as a few weeks, and is
subject to strong variations accross the SN class
(SNe 1999br or 2005cs, which boast a strong H$\alpha$ at the recombination epoch,
have symptomatically low $^{56}$Ni yields compared to the standard Type II-Plateau SN 1999em).

The time-dependence effects we observe just one week after explosion in the outer,
fully ionized, SN ejecta also support the idea that the energy gain drawn out of recombining
ions is not a fundamental driver. Time-dependence effects can lead to over-ionization
for He, C, N, O, and metals, under fully-ionized hydrogen conditions. We thus surmise
that time dependence should operate, with a magnitude to be determined, in SN ejecta
of all types, primarily because they all combine the properties of fast expansion and low density.

We have investigated the potential impact of time dependence on the correction factors
used in the Expanding Photosphere Method, or, essentially, whether the influence
on the electron density generates a systematic shift in the magnitude of flux dilution.
We find no sizable and systematic deviation from the correction factors obtained, at the same color
temperature, with steady-state CMFGEN models. This supports our use of steady-state
CMFGEN models for distance determinations based on early-time Type II SN observations
(Dessart \& Hillier 2006a; Dessart et al. 2007). However, the good agreement at late times
between time-dependent CMFGEN models and late time photospheric-phase observations
motivates the extension of the time baseline to include the recombination epoch,
thus allowing for multi-epoch observations that cover the entire Plateau phase,
thereby reducing the errors on the inferred distance.

For the modeling of the longer evolution of Type II SN ejecta and their radiation, a time-dependent
approach is warranted. To achieve a higher level of consistency, CMFGEN is under
developement to follow as well the time-dependent evolution of the radiation field,
together with options for chemical stratification and energy deposition from
radioactive isotopes. This versatility will allow us to study SN ejecta of any type
and over months after explosion.

\section*{Acknowledgments}

We thank Andrea Pastorello, Doug Leonard, and Mario Hamuy for providing optical spectra
of SN 1999br and SN 1999em. L.D. acknowledges support for this work from the
Scientific Discovery through Advanced Computing (SciDAC) program of the DOE,
under grant numbers DE-FC02-01ER41184 and DE-FC02-06ER41452, and from the
NSF under grant number AST-0504947.

\appendix

\section{}

We illustrate how the time-dependent terms can influence the He\,{\sc i} ionization in SN models.
We choose a model with a Hubble age of 6.75 days, and choose a single depth in the
atmosphere for which helium is still appreciably ionized in the time-dependent model,
but neutral when such terms are not included.

In Table~\ref{tab_term_comp} we list the various processes, and their rates, which influence the helium ionization balance. From this table we see the following: The most important process populating level
1s$^2$ (the ground state) is
decays from the 2p $^1$P$^o$ state through the optically thick resonance transition, followed
by the slight imbalance between photoionization from level 1s$^2$,  and direct
recombination to level 1s$^2$. Many other processes
also play a role, including collisional de-excitation from the 2s \& 2p levels, decay from the
2p $^3$P$^o$ state, and decays from other members of the np $^1$P$^o$ ($n>2$) series.
As expected, the sum of all these decay processes is the same as Dn(He[n=1])/Dt. Note that
Dn(He(n=1))/Dt$=1.2 \times 10^4$ cm$^{-3}$ s$^{-1}$ if helium was to change from fully ionized, to fully neutral, over one day.
Although not directly apparent from the table, most of the recombinations to excited states of neutral
helium are balanced by ionizations from the 2s and 2p states.

\begin{table}
  \centering
  \begin{minipage}{9cm}
  \caption{Term comparison. Note that PR(n=1)/RR is the photoionization/recombination rate
from/to level 1.\label{tab_term_comp}}
  \begin{tabular}{lrl}
  \hline
Process &  Term &  \\
  \hline
R                                                             &  $4.42 \times 10^{14}$           & cm    \\
v                                                             &  $7574$                          & \kms \\
He$^+$ density                                                &  $1.06 \times 10^9$              & cm$^{-3}$ \\
He(1s$^2$ $^1$S) density                                      &  $7.08 \times 10^8$              & cm$^{-3}$ \\
Electron density                                              &   $9.92 \times 10^9$             & cm$^{-3}$ \\
$t_{\hbox{exp}}$                                              &   6.75                           & days \\
$t_{\hbox{rec}}$(H)                                           &   0.0045                         & days \\
PR        (n=1)                                               &  $1.8012 \times 10^6$            & cm$^{-3}$ s$^{-1}$ \\
                                                                                               &      \\
RR-PR (n=1)                                                   &          $8.0 \times 10^2$       & cm$^{-3}$ s$^{-1}$ \\
2p $^1$P$^o \rightarrow$ 1s$^2$ $^1$S                         &          $2.15 \times 10^3$      & cm$^{-3}$ s$^{-1}$ \\
2p $^3$P$^o  \rightarrow$ 1s$^2$ $^1$S                        &          $4.49 \times 10^2$      & cm$^{-3}$ s$^{-1}$ \\
2s $^1$S  $\rightarrow$ 1s$^2$ $^1$S                          &          $6.4 \times 10^1$       & cm$^{-3}$ s$^{-1}$ \\
$\sum_{n>2}$ np $^1$P$^o \rightarrow$ 1s$^2$ $^1$S            &          $3.00 \times 10^2$      & cm$^{-3}$ s$^{-1}$ \\
$\sum$ 2s, 2p states $\rightarrow$  1s$^2$ $^1$S  (col)      &          $3.62 \times 10^2$      & cm$^{-3}$ s$^{-1}$ \\
Dn(He(n=1))/Dt                                                &   $4.11 \times 10^3 $            & cm$^{-3}$ s$^{-1}$ \\
\hline
\end{tabular}
\end{minipage}
\end{table}


\label{lastpage}

\end{document}